\documentclass[preprint]{aastex62}
\usepackage{booktabs}
\usepackage{graphicx}
\usepackage{blindtext}

\newcommand{\msun}{{\rm \ M_\odot}}
\newcommand{\bdv}[1]{\mbox{\boldmath$#1$}}

\newcommand\ltsima{$\; \buildrel <\over\sim \;$}
\newcommand\simlt{\lower.5ex\hbox{\ltsima}}
\newcommand\gtsima{$\; \buildrel >\over\sim \;$}
\newcommand\simgt{\lower.5ex\hbox{\gtsima}}
\def\au{{\rm au}} 
 
\def\kms{{\rm km}\,{\rm s}^{-1}}
\def\masyr{{\rm mas}\,{\rm yr}^{-1}}
\def\kpc{{\rm kpc}}
\def\mas{{\rm mas}}
\def\sat{{\rm sat}}
\def\muas{\mu{\rm as}}

\def\rel{{\rm rel}}

\def\e{{\rm E}}
\def\bpi{{\bdv\pi}}
\def\bmu{{\bdv\mu}}

\def\bv{{\bf v}}

\NewPageAfterKeywords

\begin{document}
\title{OGLE-2017-BLG-0406: {\it Spitzer} Microlens Parallax Reveals Saturn-mass Planet orbiting M-dwarf Host in the Inner Galactic Disk}

\author{Yuki Hirao}
\affiliation{Depertment of Earth and Space Science, Graduate School of Science, Osaka University, 1-1 Machikaneyama, Toyonaka, Osaka 560-0043, Japan}
\affiliation{Laboratory for Exoplanets and Stellar Astrophysics, NASA / Goddard Space Flight Center, Greenbelt, MD 20771, USA}
\affiliation{Department of Astronomy, University of Maryland, College Park, MD 20742, USA}
\author{David P. Bennett}
\affiliation{Laboratory for Exoplanets and Stellar Astrophysics, NASA / Goddard Space Flight Center, Greenbelt, MD 20771, USA}
\affiliation{Department of Astronomy, University of Maryland, College Park, MD 20742, USA}
\author{Yoon-Hyun Ryu}
\affiliation{Korea Astronomy and Space Science Institute, Daejon 34055, Republic of Korea}
\author{Naoki Koshimoto}
\affiliation{Department of Astronomy, Graduate School of Science, The University of Tokyo, 7-3-1 Hongo, Bunkyo-ku, Tokyo 113-0033, Japan}
\author{Andrzej Udalski}
\affiliation{Astronomical Observatory, University of Warsaw, Al. Ujazdowskie 4, 00-478 Warszawa, Poland}
\author{Jennifer C. Yee}
\affiliation{Center for Astrophysics $|$ Harvard \& Smithsonian, 60 Garden St.,Cambridge, MA 02138, USA}
\author{Takahiro Sumi}
\affiliation{Depertment of Earth and Space Science, Graduate School of Science, Osaka University, 1-1 Machikaneyama, Toyonaka, Osaka 560-0043, Japan}
\author{Ian A. Bond}
\affiliation{Institute of Information and Mathematical Sciences, Massey University, Private Bag 102-904, North Shore Mail Centre, Auckland, New Zealand}
\author{Yossi Shvartzvald}
\affiliation{Department of Particle Physics and Astrophysics, Weizmann Institute of Science, Rehovot 76100, Israel}

\collaboration{and}

\author{Fumio Abe}
\affiliation{Institute for Space-Earth Environmental Research, Nagoya University, Nagoya 464-8601, Japan}
\author{Richard K. Barry}
\affiliation{Astrophysics Science Division, NASA/Goddard Space Flight Center, Greenbelt, MD20771, USA}
\author{Aparna Bhattacharya}
\affiliation{Laboratory for Exoplanets and Stellar Astrophysics, NASA / Goddard Space Flight Center, Greenbelt, MD 20771, USA}
\affiliation{Department of Astronomy, University of Maryland, College Park, MD 20742, USA}
\author{Martin Donachie}
\affiliation{Department of Physics, University of Auckland, Private Bag 92019, Auckland, New Zealand}
\author{Akihiko Fukui}
\affiliation{Department of Earth and Planetary Science, Graduate School of Science, The University of Tokyo, 7-3-1 Hongo, Bunkyo-ku, Tokyo 113-0033, Japan}
\affiliation{Instituto de Astrof\'isica de Canarias, V\'ia L\'actea s/n, E-38205 La Laguna, Tenerife, Spain}
\author{Yoshitaka Itow}
\affiliation{Institute for Space-Earth Environmental Research, Nagoya University, Nagoya 464-8601, Japan}
\author{Iona Kondo}
\affiliation{Depertment of Earth and Space Science, Graduate School of Science, Osaka University, 1-1 Machikaneyama, Toyonaka, Osaka 560-0043, Japan}
\author{Man Cheung Alex Li}
\affiliation{Department of Physics, University of Auckland, Private Bag 92019, Auckland, New Zealand}
\author{Yutaka Matsubara}
\affiliation{Institute for Space-Earth Environmental Research, Nagoya University, Nagoya 464-8601, Japan}
\author{Taro Matsuo}
\affiliation{Depertment of Earth and Space Science, Graduate School of Science, Osaka University, 1-1 Machikaneyama, Toyonaka, Osaka 560-0043, Japan}
\author{Shota Miyazaki}
\affiliation{Depertment of Earth and Space Science, Graduate School of Science, Osaka University, 1-1 Machikaneyama, Toyonaka, Osaka 560-0043, Japan}
\author{Yasushi Muraki}
\affiliation{Institute for Space-Earth Environmental Research, Nagoya University, Nagoya 464-8601, Japan}
\author{Masayuki Nagakane}
\affiliation{Depertment of Earth and Space Science, Graduate School of Science, Osaka University, 1-1 Machikaneyama, Toyonaka, Osaka 560-0043, Japan}
\author{Cl\'ement Ranc}
\affiliation{Laboratory for Exoplanets and Stellar Astrophysics, NASA / Goddard Space Flight Center, Greenbelt, MD 20771, USA}
\author{Nicholas J. Rattenbury}
\affiliation{Department of Physics, University of Auckland, Private Bag 92019, Auckland, New Zealand}
\author{Haruno Suematsu}
\affiliation{Depertment of Earth and Space Science, Graduate School of Science, Osaka University, 1-1 Machikaneyama, Toyonaka, Osaka 560-0043, Japan}
\author{Hiroshi Shibai}
\affiliation{Depertment of Earth and Space Science, Graduate School of Science, Osaka University, 1-1 Machikaneyama, Toyonaka, Osaka 560-0043, Japan}
\author{Daisuke Suzuki}
\affiliation{Depertment of Earth and Space Science, Graduate School of Science, Osaka University, 1-1 Machikaneyama, Toyonaka, Osaka 560-0043, Japan}
\author{Paul J. Tristram}
\affiliation{University of Canterbury Mt. John Observatory, P.O. Box 56, Lake Tekapo 8770, New Zealand}
\author{Atsunori Yonehara}
\affiliation{Department of Physics, Faculty of Science, Kyoto Sangyo University, Kyoto 603-8555, Japan}

\collaboration{(the MOA Collaboration)}

\author{J. Skowron}
\affiliation{Astronomical Observatory, University of Warsaw, Al. Ujazdowskie 4, 00-478 Warszawa, Poland}
\author{R. Poleski}
\affiliation{Astronomical Observatory, University of Warsaw, Al. Ujazdowskie 4, 00-478 Warszawa, Poland}
\affiliation{Department of Astronomy, Ohio State University, 140 West 18th Avenue, Columbus, OH 43210, USA}
\author{P. Mr\'oz}
\affiliation{Division of Physics, Mathematics, and Astronomy, California Institute of Technology, Pasadena, CA 91125, USA}
\author{M.K. Szyma\'nski}
\affiliation{Astronomical Observatory, University of Warsaw, Al. Ujazdowskie 4, 00-478 Warszawa, Poland}
\author{I. Soszy\'nski}
\affiliation{Astronomical Observatory, University of Warsaw, Al. Ujazdowskie 4, 00-478 Warszawa, Poland}
\author{S. Koz{\l}owski}
\affiliation{Astronomical Observatory, University of Warsaw, Al. Ujazdowskie 4, 00-478 Warszawa, Poland}
\author{P. Pietrukowicz}
\affiliation{Astronomical Observatory, University of Warsaw, Al. Ujazdowskie 4, 00-478 Warszawa, Poland}
\author{K. Ulaczyk}
\affiliation{Astronomical Observatory, University of Warsaw, Al. Ujazdowskie 4, 00-478 Warszawa, Poland}
\author{K. Rybicki}
\affiliation{Astronomical Observatory, University of Warsaw, Al. Ujazdowskie 4, 00-478 Warszawa, Poland}
\author{P. Iwanek}
\affiliation{Astronomical Observatory, University of Warsaw, Al. Ujazdowskie 4, 00-478 Warszawa, Poland}

\collaboration{(the OGLE Collaboration)}

\author{Michael D.Albrow}
\affiliation{University of Canterbury, Department of Physics and
Astronomy, Private Bag 4800, Christchurch 8020, New Zealand}
\author{Sun-Ju Chung}
\affiliation{Korea Astronomy and Space Science Institute, Daejon 34055, Republic of Korea}
\affiliation{Korea University of Science and Technology, Daejeon 34113, Republic of Korea}
\author{Andrew Gould}
\affiliation{Korea Astronomy and Space Science Institute, Daejon 34055, Republic of Korea}
\affiliation{Max-Planck-Institute for Astronomy, K\"{o}nigstuhl 17, 69117 Heidelberg, Germany}
\affiliation{Department of Astronomy, Ohio State University, 140 W. 18th Ave., Columbus, OH 43210, USA}
\author{Cheongho Han}
\affiliation{Department of Physics, Chungbuk National University, Cheongju 28644, Republic of Korea}
\author{Kyu-Ha Hwang}
\affiliation{Korea Astronomy and Space Science Institute, Daejon 34055, Republic of Korea}
\author{Youn Kil Jung}
\affiliation{Korea Astronomy and Space Science Institute, Daejon 34055, Republic of Korea}
\author{In-Gu Shin}
\affiliation{Korea Astronomy and Space Science Institute, Daejon 34055, Republic of Korea}
\author{Weicheng Zang}
\affiliation{Physics Department and Tsinghua Centre for Astrophysics, Tsinghua University, Beijing 100084, China}
\author{Sang-Mok Cha}
\affiliation{Korea Astronomy and Space Science Institute, Daejon 34055, Republic of Korea}
\affiliation{School of Space Research, Kyung Hee University, Yongin, Kyeonggi 17104, Republic of Korea}
\author{Dong-Jin Kim}
\affiliation{Korea Astronomy and Space Science Institute, Daejon 34055, Republic of Korea}
\author{Hyoun-Woo Kim}
\affiliation{Korea Astronomy and Space Science Institute, Daejon 34055, Republic of Korea}
\author{Seung-Lee Kim}
\affiliation{Korea Astronomy and Space Science Institute, Daejon 34055, Republic of Korea}
\affiliation{Korea University of Science and Technology, Daejeon 34113, Republic of Korea}
\author{Chung-Uk Lee}
\affiliation{Korea Astronomy and Space Science Institute, Daejon 34055, Republic of Korea}
\author{Dong-Joo Lee}
\affiliation{Korea Astronomy and Space Science Institute, Daejon 34055, Republic of Korea}
\author{Yongseok Lee}
\affiliation{Korea Astronomy and Space Science Institute, Daejon 34055, Republic of Korea}
\affiliation{School of Space Research, Kyung Hee University, Yongin, Kyeonggi 17104, Republic of Korea}
\author{Byeong-Gon Park}
\affiliation{Korea Astronomy and Space Science Institute, Daejon 34055, Republic of Korea}
\affiliation{Korea University of Science and Technology, Daejeon 34113, Republic of Korea}
\author{Richard W. Pogge}
\affiliation{Department of Astronomy, Ohio State University, 140 W. 18th Ave., Columbus, OH 43210, USA}

\collaboration{(the KMTNet Collaboration)}

\author{Charles A. Beichman}
\affiliation{IPAC, Mail Code 100-22, Caltech, 1200 E. California Blvd., Pasadena, CA 91125, USA}
\author{Geoffery Bryden}
\affiliation{Jet Propulsion Laboratory, California Institute of Technology, 4800 Oak Grove Drive, Pasadena, CA 91109, USA}
\author{Sebastiano Calchi Novati}
\affiliation{IPAC, Mail Code 100-22, Caltech, 1200 E. California Blvd., Pasadena, CA 91125, USA}
\author{Sean Carey}
\affiliation{IPAC, Mail Code 100-22, Caltech, 1200 E. California Blvd., Pasadena, CA 91125, USA}
\author{B. Scott Gaudi}
\affiliation{Department of Astronomy, Ohio State University, 140 W. 18th Ave., Columbus, OH  43210, USA}
\author{Calen B. Henderson}
\affiliation{IPAC, Mail Code 100-22, Caltech, 1200 E. California Blvd., Pasadena, CA 91125, USA}
\author{Wei Zhu}
\affiliation{Canadian Institute for Theoretical Astrophysics, University of Toronto, 60 St George Street, Toronto, ON M5S 3H8, Canada}

\collaboration{(the Spitzer team)}

\author{Etienne Bachelet}
\affiliation{Las Cumbres Observatory, 6740 Cortona Drive, suite 102, Goleta, CA 93117, USA}
\author{Greg Bolt}
\affiliation{Craigie Observatory, Western Australia, Australia}
\author{Grant Christie}
\affiliation{Auckland Observatory, Auckland, New Zealand}
\author{Markus Hundertmark}
\affiliation{Astronomisches Rechen-Institut, Zentrum f{\"u}r Astronomie der Universit{\"a}t Heidelberg (ZAH), 69120 Heidelberg, Germany}
\author{Tim Natusch}
\affiliation{Auckland Observatory, Auckland, New Zealand ; Institute for Radio Astronomy and Space Research (IRASR), AUT University, Auckland, New Zealand}
\author{Dan Maoz}
\affiliation{School of Physics and Astronomy, Tel-Aviv University, Tel-Aviv 6997801, Israel}
\author{Jennie McCormick}
\affiliation{Farm Cove Observatory, Centre for Backyard Astrophysics, Pakuranga, Auckland, New Zealand}
\author{Rachel  A. Street}
\affiliation{Las Cumbres Observatory, 6740 Cortona Drive, suite 102, Goleta, CA 93117, USA}
\author{Thiam-Guan Tan}
\affiliation{Perth Exoplanet Survey Telescope, Perth, Australia}
\author{Yiannis Tsapras}
\affiliation{Astronomisches Rechen-Institut, Zentrum f{\"u}r Astronomie der Universit{\"a}t Heidelberg (ZAH), 69120 Heidelberg, Germany}

\collaboration{(the LCO and $\mu$FUN Follow-up Teams)}

\author{U. G. J{\o}rgensen}
\affiliation{Niels Bohr Institute \& Centre for Star and Planet Formation, University of Copenhagen {\O}ster Voldgade 5, 1350 - Copenhagen, Denmark}
\author{M. Dominik}
\affiliation{Centre for Exoplanet Science, SUPA School of Physics \& Astronomy, University of St Andrews, North Haugh, St Andrews, KY16 9SS}
\author{V. Bozza}
\affiliation{Dipartimento di Fisica "E.R. Caianiello", Universit{\`a} di Salerno, Via Giovanni Paolo II 132, 84084, Fisciano, Italy}
\affiliation{Istituto Nazionale di Fisica Nucleare, Sezione di Napoli, Napoli, Italy}
\author{J. Skottfelt}
\affiliation{Centre for Electronic Imaging, Department of Physical Sciences, The Open University, Milton Keynes, MK7 6AA, UK}
\author{C. Snodgrass}
\affiliation{Institute for Astronomy, University of Edinburgh, Royal Observatory, Edinburgh EH9 3HJ, UK}
\author{S. Ciceri}
\affiliation{Department of Astronomy, Stockholm University, AlbaNova University Centre, 106 91Stockholm, Sweden}
\author{R. Figuera Jaimes}
\affiliation{Centre for Exoplanet Science, SUPA School of Physics \& Astronomy, University of St Andrews, North Haugh, St Andrews, KY16 9SS}
\affiliation{Physics Department and Tsinghua Centre for Astrophysics, Tsinghua University, Beijing 100084, China}
\author{D. F. Evans}
\affiliation{Astrophysics Group, Keele University, Staffordshire, ST5 5BG, UK}
\author{N. Peixinho}
\affiliation{Centro de Astronom{\'{\i}}a (CITEVA),  Universidad de Antofagasta, Avda. U. de Antofagasta 02800, Antofagasta, Chile}
\author{T. C. Hinse}
\affiliation{Chungnam National University, Department of Astronomy and Space Science, 34134 Daejeon, Republic of Korea}
\author{M. J. Burgdorf}
\affiliation{Universit{\"a}t Hamburg, Faculty of Mathematics, Informatics and Natural Sciences, Department of Earth Sciences, Meteorological Institute, Bundesstra\ss{}e 55, 20146 Hamburg, Germany}
\author{J. Southworth}
\affiliation{Astrophysics Group, Keele University, Staffordshire, ST5 5BG, UK}
\author{S. Rahvar}
\affiliation{Department of Physics, Sharif University of Technology, PO Box 11155-9161 Tehran, Iran}
\author{S. Sajadian}
\affiliation{Department of Physics, Isfahan University of Technology, Isfahan, Iran}
\author{M. Rabus}
\affiliation{Instituto de Astrof\'isica, Pontificia Universidad Cat\'olica de Chile, Av. Vicu\~na Mackenna 4860, 7820436 Macul, Santiago, Chile}
\author{C. von Essen}
\affiliation{Stellar Astrophysics Centre, Department of Physics and Astronomy, Aarhus University, Ny Munkegade 120, 8000 Aarhus C, Denmark}
\author{Y. I. Fujii}
\affiliation{Niels Bohr Institute \& Centre for Star and Planet Formation, University of Copenhagen {\O}ster Voldgade 5, 1350 - Copenhagen, Denmark}
\affiliation{Institute for Advanced Research, Nagoya University, Furo-cho, Chikusa-ku, Nagoya, 464-8601, Japan}
\author{J. Campbell-White}
\affiliation{Centre for Astrophysics \& Planetary Science, The University of Kent, Canterbury CT2 7NH, UK}
\author{S. Lowry}
\affiliation{Centre for Astrophysics \& Planetary Science, The University of Kent, Canterbury CT2 7NH, UK}
\author{C. Helling}
\affiliation{Centre for Exoplanet Science, SUPA School of Physics \& Astronomy, University of St Andrews, North Haugh, St Andrews, KY16 9SS}
\author{L. Mancini}
\affiliation{Dipartimento di Fisica, Universit\'a di Roma Tor Vergata, Via della Ricerca Scientifica 1, 00133 Roma, Italy}
\affiliation{Max Planck Institute for Astronomy, K{\"o}nigstuhl 17, 69117 Heidelberg, Germany}
\affiliation{INAF -- Astrophysical Observatory of Turin, Via Osservatorio 20, 10025 Pino Torinese, Italy}
\affiliation{International Institute for Advanced Scientific Studies (IIASS), Via G. Pellegrino 19, 84019 Vietri sul Mare (SA), Italy}
\author{L. Haikala}
\affiliation{Universidad de Atacama, Copiapo, Chile}

\collaboration{(the MindSTEp Collaboration)}

\author{Ryo Kandori}
\affiliation{Astrobiology Center of NINS, 2-21-1, Osawa, Mitaka, Tokyo 181-8588, Japan}

\collaboration{(the IRSF team)}

\begin{abstract}
We report the discovery and analysis of the planetary microlensing
event OGLE-2017-BLG-0406, which was observed both from the ground and
by the {\it Spitzer} satellite in a solar orbit. At high magnification,
the anomaly in the light curve was densely observed by ground-based-survey 
and follow-up groups, and it was found to be explained by a planetary lens with a 
planet/host mass ratio of $q=7.0\times 10^{-4}$ from the light-curve
modeling. The ground-only and {\it Spitzer}-``only'' data each
provide very strong one-dimensional (1-D) constraints on the 2-D
microlens parallax vector $\bpi_\e$. When combined, these yield
a precise measurement of $\bpi_\e$, and so of the masses of the
host $M_{\rm host}=0.56\pm0.07\,M_\odot$ and 
planet $M_{\rm planet} = 0.41\pm 0.05\,M_{\rm Jup}$. The system lies at a distance 
$D_{\rm L}=5.2\pm 0.5\,\kpc$ from the Sun toward the Galactic bulge, and 
the host is more likely to be a disk population star according to the kinematics of the lens.  
The projected separation of the planet from the host is
$a_\perp = 3.5\pm 0.3\, \rm au$, i.e., just over twice the snow line.
The Galactic-disk kinematics are established in part from a precise measurement
of the source proper motion based on OGLE-IV data.  By contrast,
the {\it Gaia} proper-motion measurement of the source suffers from a 
catastrophic $10\,\sigma$ error.
\end{abstract}


\section{INTRODUCTION}
Gravitational microlensing has a unique strength in its sensitivity to planets with masses as low as Earth-mass \citep{ben96} just beyond the snow line \citep{gl92}, where the core accretion theory of planetary formation predicts the most efficient planet formation \citep{ida05}. Because it does not rely on the light from the host star, microlensing can detect the planets orbiting around faint stars like M-dwarfs and brown dwarfs, and can even detect free-floating planets \citep{sum11,mro17,ob161540,ob170560,mro20}. Microlensing can also detect planets in the Galactic bulge because microlensing events can be caused by stars at any distance between Earth and the Galactic bulge, where most of the stars that act as sources lie. This is complementary to other planet detection techniques such as the radial velocity \citep{but06} and transit \citep{bor11} methods, which are most sensitive to planets in short period orbits. Therefore, the microlensing method is essential for the complete demographics census of Galactic planetary systems \citep{gau12, tsa18}. 
\\
\par
Several statistical studies based on the discovered microlensing planets have been conducted and revealed the planet occurrence rates beyond the snow line \citep{gou10,sum10,cas12,shv16} and the possible paucity of planets in the Galactic bulge \citep{pen16}. One of the most important microlensing statistical results is that of \citet{suz16}, who found a clear break and likely peak in the planet-host mass ratio function at a mass ratio of $q \sim 10^{-4}$ using 30 exoplanets detected by microlensing. This peak was confirmed by \citet{ob171434} and \citet{jun18}, who determined that the peak occured at a mass ratio of $q \approx 6\times 10^{-5}$.
A comparison of the \citet{suz16} results to population synthesis models based on the core accretion theory \citep{suz18} reveals a discrepancy between the smooth mass ratio distribution for the microlens planets and the predicted deficit of planets with mass ratios lying in the range of $10^{-4} < q < 4\times 10^{-4}$. This predicted gap in the mass ratio distribution \citep{ida04} is due to the runaway gas accretion process \citep{pol96,lissauer09}, which has long been considered a fundamental aspect of the core accretion theory. So, the microlensing results seem to imply that a major change in the theory is needed. In fact, recent  three-dimensional high resolution numerical calculations (\citealt{szu14};  Szul{\'a}gyi et al. in preparation) indicate  that runaway gas accretion often halted or decreased due to the circumplanetary disk formation and suggest that earlier, lower resolution three-dimensional calculations had numerical artifacts that favored the runaway gas accretion scenario.
Comparison of the population synthesis results to ALMA protoplanetary disk observations also support this conclusion 
\citep{nay19}.
\\
\par
Microlensing light curve models provide the lens planet-host mass ratios, but they do not usually provide the lens mass and distance. To measure the properties of lens systems, one needs additional observables that yields mass-distance relation of the lens systems such as the angular Einstein ring radius $\theta_{\rm E}$ and the microlens parallax $\pi_{\rm E}$. The measurement of the $\theta_{\rm E}$ or $\pi_{\rm E}$ values yields the following mass-distance relations,
\begin{equation}
M_L = {c^2\over 4G} \theta_E^2 {D_{\rm S} D_{\rm L}\over D_{\rm S} - D_{\rm L}} 
       =  {c^2\over 4G}{ {\rm au}\over{\pi_E}^2}{D_{\rm S} - D_{\rm L}\over D_{\rm S}  D_{\rm L}}  \ ,
\label{eq-mDl}
\end{equation}
where $D_{\rm L}$ is the lens distance and the source distance, $D_{\rm S}$, is known (approximately). If the apparent $K$-band magnitude of the lens star $K_{L,\rm meas}$ is measured, then we have a mass-distance relation given by $K_{L,\rm meas} = 5\log_{10}(D_{\rm L}/10{\rm pc}) + A_K(D_{\rm L}) + K_{\rm abs, meas}(M_L)$, where $K_{\rm abs, meas}(M_L)$ is a $K$-band mass-luminosity relation and $A_K(D_{\rm L})$ is a model of the extinction in the foreground of the lens star. Measurements of the lens brightness in other passbands yield independent mass-distance relations. Combining any two of these mass-distance relations will yield the lens mass $M_{\rm L}$ and distance $D_{\rm L}$. The most elegant solution is obtained if both the angular Einstein radius, $\theta_{\rm E}$, and the microlensing parallax, $\pi_{\rm E}$, are measured because this distance dependence cancels, 
enabling unique determinations of $M_{\rm L}$ and $D_{\rm L}$ by the following relations, 
\begin{equation}
M_{\rm L} = \frac{\theta_{\rm E}}{\kappa \pi_{\rm E}}; \ D_{\rm L} = \frac{\rm au}{\pi_{\rm E} \theta_{\rm E} + \pi_{\rm S}},
\end{equation}
where $\kappa = 4G/(c^2 {\rm au}) = 8.1439 \ {\rm mas}/M_{\odot}$ and $\pi_{\rm S} = 1{\rm au}/D_{\rm S}$ \citep{gould92,gou00}. 
For binary events, $\theta_{\rm E}$ can be routinely measured by measuring the source radius crossing time, $t_*$, 
provided that the source crosses a caustic curve or closely approaches to a caustic cusp. 
This gives $\theta_{\rm E} = \theta_* t_{\rm E}/t_*$,
where $ \theta_*$ is the angular radius of the source, which can be determined from the light curve model
values for the source brightness and color \citep{alb98,yoo04}.
\\
\par
It can be challenging to make the measurements necessary for the other mass-distance relations besides the 
$\theta_{\rm E}$ relation. Detecting the host star is nearly impossible for bright source stars, and a unique identification 
of the host star can be difficult if the source star is bright (i.e.,\ a giant star) or if the relative lens-source proper motion is 
not big enough to resolve the lens and source 
\citep{bha17,kos17,kos19keck}. Because the lens-source separation increase as time passes after an event, there are an increasing
number of planetary events with mass measurements from host star brightness measurements 
\citep{bat15,ben06,ben15,ben20,bha18,van19}, and this is the method that is expected to make most of the exoplanet
mass measurements for WFIRST \citep{ben02,ben07,spe15}.
\\
\par
The microlensing parallax effect has traditionally been measured due to the effects of the orbital motion of Earth.
\citet{dong-ogle71} made the first such measurement on
OGLE-2005-BLG-071 (only the second planet detected by microlensing
\citep{uda05}), which was made possible in part by the exceptionally
 large parallax\footnote{\citet{ben20} confirmed
this first planet-event parallax measurement and found a $2\,\sigma$
correction, using high-resolution imaging}.
However, in general, this annual parallax effect can only be measured
 for a subset of planetary microlensing events: events that have long durations, like
OGLE-2006-BLG-109 \citep{gau08,ben10b} and OGLE-2007-BLG-349 \citep{ben16}, have bright source stars and
moderately long durations, like MOA-2009-BLG-266 \citep{mur11} and OGLE-2012-BLG-0265 \citep{sko15}, or have very special
lens-source geometries, such as MOA-2013-BLG-605 \citep{sumi16} and OGLE-2013-BLG-0341 \citep{gou14}.
 \\
\par
However, $\pi_{\rm E}$ can also be measured by simultaneously observing lensing events from two well-separated ($\sim$ au) 
observatories \citep{refsdal66}. Since 2014, almost 1000 events including both single and binary events were simultaneously 
observed from the ground and the {\it Spitzer} Space Telescope \citep{yee15,zhu17}. {\it Spitzer} observations helped to 
determine the distance to the lens for over a hundred of those events. To date, ten 
planetary events were observed by {\it Spitzer}. Seven of these are located in the Galactic disk:
OGLE-2014-BLG-0124 \citep{ob140124,bea18}, OGLE-2015-BLG-0966 \citep{ob150966},
OGLE-2017-BLG-1140 \citep{ob171140}, OGLE-2016-BLG-1067 \citep{ob161067}, OGLE-2016-BLG-1195 \citep{ob161195a,ob161195b} KMT-2018-BLG-0029 \citep{kb180029} and Kojima-1 \citep{nucita18,fukui19,kojima2}. 
The lens systems for event OGLE-2016-BLG-1190 \citep{ob161190}, OGLE-2018-BLG-0596 \citep{ob180596} and  OGLE-2018-BLG-0799 \citep{ob180799} are reported to be in the Galactic bulge. While observations from {\it Spitzer} make it easier to measure the small $\pi_{\rm E}$ values for bulge lens systems, this ability is undermined by the requirement that events should be discovered at least $\sim 1$ week before {\it Spitzer} observations can be requested (Figure 1 from \citet{ob140124}). 
This combined with the limited 40-day {\it Spitzer} observing window for bulge events leads to incomplete light curves, which can make parallax measurements difficult.
\\
\par
In this paper, we report the discovery and the analysis of the planetary microlensing event OGLE-2017-BLG-0406, which was observed both from the ground and in space using the {\it Spitzer} telescope. The anomaly in the light curve was well covered by ground-based observations. The additional {\it Spitzer} data constrained the parallax parameters, hence the mass and the distance of the lens systems. We describe the ground-based and space-based observations in Section~\ref{sec:obs} and the data reductions in Section~\ref{sec:data}. In Section~\ref{sec:anal}, we describe our light curve modeling conducted for the ground-based data. We present our {\it Spitzer} parallax analysis in Section~\ref{sec:spitzpar}. In Section~\ref{sec:cmd} to Section~\ref{sec:physical}, we present the determinations of source properties and lens properties. Finally, we discuss and summarize the results in Section~\ref{sec:discuss}. 
\\
\par

\section{OBSERVATIONS}
\label{sec:obs}
\subsection{Ground Based Observation}
The microlensing event OGLE-2017-BLG-0406 was first discovered on March 27 (HJD$^\prime$ = HJD-2450000 = 7839) by the Optical Gravitational Lensing Experiment (OGLE) collaboration at (R.A., decl)(J2000) = ($17^{h}55^{m}59^{s}.92$,$-29^{\circ}51'47''.3$) or $(l,b) = (0^{\circ}.3601,-2^{\circ}.4164)$ in Galactic coordinates and alerted by the OGLE Early Warning System \citep{uda03}. The event lies in the OGLE-IV field BLG506, and the observations were conducted at the cadence of once per hour by using the 1.3m Warsaw telescope located at Las Campanas Observatory in Chile, equipped with a 1.4 $\rm deg^2$ field-of-view CCD camera. The Microlensing Observations in Astrophysics (MOA) group independently discovered this event on May 5 (HJD$^\prime$ = 7879) by using the MOA alert system \citep{bon01} and identified it as MOA-2017-BLG-233. MOA observed this event with 15 minutes cadence by using MOA-II telescope at Mt. John University Observatory in New Zealand, equipped with 2.2 $\rm deg^2$ field of view camera MOA-camIII \citep{sak08}. Most observations were conducted in the customized MOA-Red wide band, which is the sum of the standard Cousins {\sl R} and {\sl I} bands with occasional observations in the Johnson {\sl V} band. The event was also independently discovered as KMT-2017-BLG-0243 by the Korean Microlensing Network (KMTNet: \citealt{kim16}) survey using its post-season event finder \citep{kim18}. KMTNet observes toward the Galactic bulge by using three 1.6m telescopes equipped with 4 $\rm deg^2$ camera at the Cerro Tololo Inter-American Observatory in Chile (CTIO: KMT-C), the South African Astronomical Observatory in South Africa (SAAO: KMT-S) and the Siding Spring Observatory in Australia (SSO: KMT-A). Because this event was in an overlapping region between two fields (KMTNet BLG02 and BLG42), the observations were conducted at a 15 minute cadence. 
\\
\par
\begin{figure}
\plotone{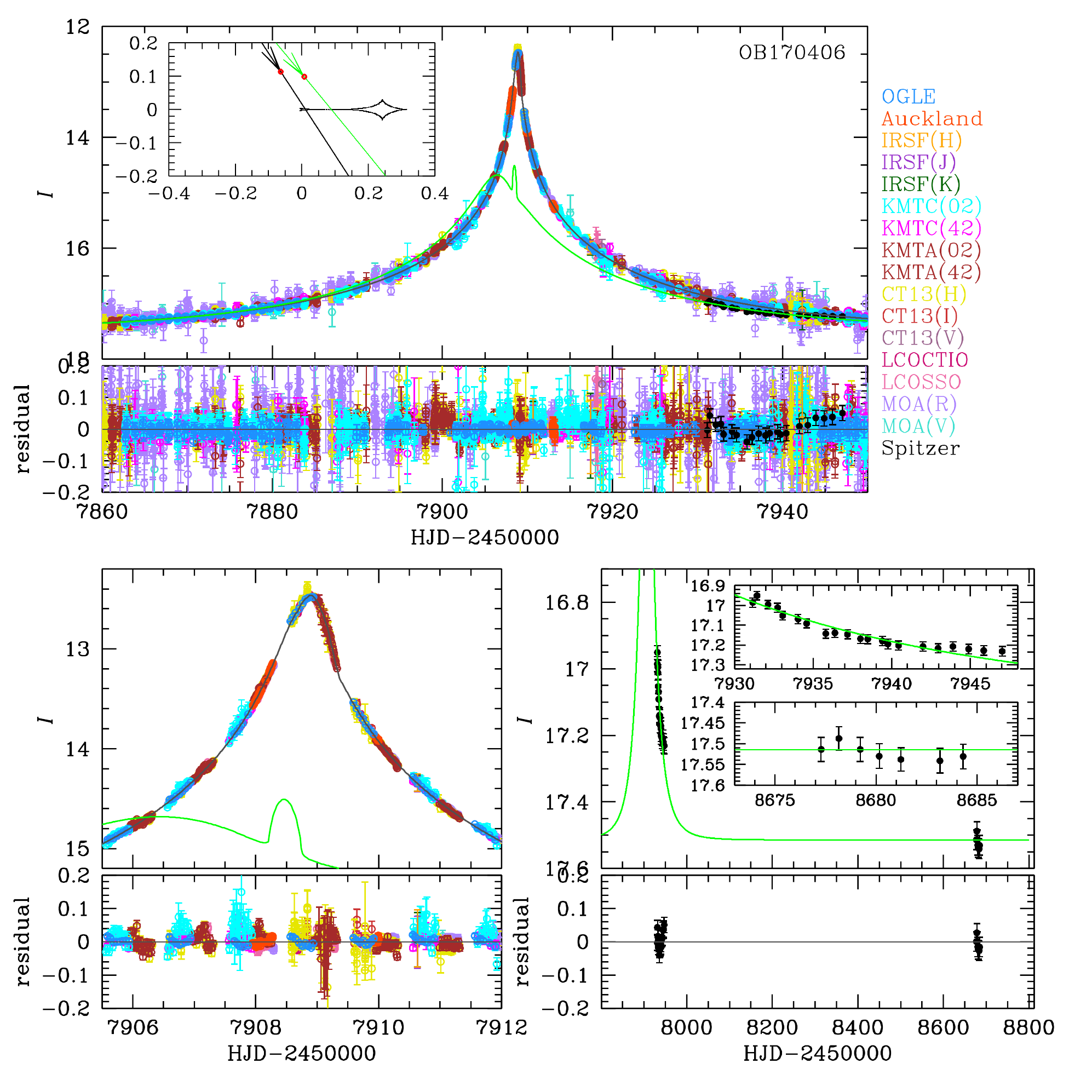}
\caption{Observed light curve of OGLE-2017-BLG-0406 and best-fit model light curve for the Wide (+,+) model. 
Data points from
different collaborations are shown with different colors. The blue
and green solid lines are the model light curves for the ground and
{\it Spitzer} observations. 
The
bottom left and right panels show a close-up of the anomaly and
{\it Spitzer} observations with the residuals from the best-fit model,
respectively. The insert on the top panel shows 
the caustic structure.  
}
\label{fig:lc}
\end{figure}

On June 2 (HJD$^\prime$=7907), the Microlensing Follow-up Network ($\mu$FUN) collaboration issued an alert that the event was peaking at a high magnification, which means that there is a high probability that the light curve will show an anomaly if the lens star hosts a planet \citep{gri98}. After the alert, $\mu$FUN, the Microlensing Network for the Detection of Small Terrestrial Exoplanet (MiNDSTEp) collaboration and Las Cumbres Observatory (LCO) global network of  telescope collaboration started high-cadence follow-up observations. $\mu$FUN used the following telescopes: the 1.3m CTIO telescope in Chile, the 0.41m Auckland telescope and the 0.36m Farm Cove telescope in New Zealand, and the 0.30m Perth Exoplanet Survey Telescope (PEST), and the 0.25m Craigie telescope in Australia. MiNDSTEp used the 1.54m Danish Telescope at La Silla Observatory in Chile. LCO used the 1.0m telescopes at CTIO in Chile and at SSO in Australia. Figure \ref{fig:lc} shows the light curve of the event. 
\\
\par

On June 4 (HJD$^\prime$ = 7909), deviations from a single lens fit were noticed just after the peak by the MOA observer. Then the first planetary model was circulated by V.~Bozza, and it was confirmed by several modelers. Because the event was very bright ($\sim$12.5 mag in {\sl I-}band), some images taken with normal exposure time by survey telescopes were saturated. 
\\
\par
We also obtained three near infrared images taken at different epochs (HJD$^\prime$ $\sim$ 7911, 7918 and 7942). The observations were made with SIRIUS, a simultaneous imager in {\sl J, H and $K_{S}$} bands, covering an area 7.7 $\times$ 7.7 $\rm arcmin^2$ with a pixel scale of $0''.45$ \citep{nag03} on the 1.4m InfraRed Survey Facility (IRSF) telescope at SAAO. The observations were conducted to measure the source color rather than for light curve modeling. The data sets are listed in Table \ref{tab1}.
\\
\par

\begin{deluxetable}{lcccc}
\tablecaption{The Data Sets Used to Model the OGLE-2017-BLG-0406 Light Curve and the Error Correction Parameters \label{tab1}}
\tablehead{
\colhead{Telescope} & \colhead{filter} & \colhead{$N_{use} / N_{obs}$} &  \colhead{$k$} & \colhead{$e_{\sl min}$} \\
}
\startdata
OGLE & {\sl I} &  6185 / 6185 &1.441 & 0.003257  \\
MOA & Red &  14611 / 14611 &1.970 & 0.003257 \\
 & {\sl V} & 392 / 392 & 1.794 & 0.003257 \\
KMT-C02 & {\sl I} & 1558 / 1558 & 2.537 & 0.003257 \\
KMT-C42 & {\sl I} & 1713 / 1713 & 1.897 & 0.003257 \\
KMT-A02 & {\sl I} & 1903 / 1903 & 2.590 & 0.003257 \\
KMT-A42 & {\sl I} & 2068 / 2068 & 2.550 & 0.003257 \\
KMT-S02 & {\sl I} & 0 / 2485 &  &  \\
KMT-S42 & {\sl I} & 0 / 2481 &  &  \\
Danish &  {\sl I} & 0 / 200 & &  \\
LCO-CTIO & {\sl I} & 296 / 296 & 1.323 & 0.003257 \\
LCO-SSO & {\sl I} & 464 / 464 & 1.644 & 0.003257 \\
Auckland & {\sl R} & 82 / 82 & 2.280 & 0.003257 \\
Craigie & clear & 0 / 677 & & \\
Farm Cove & clear & 0 / 31 & & \\
PEST & clear & 0 / 262 & &  \\
CTIO & {\sl I} & 21 / 21 & 0.860 & 0.003257 \\
 & {\sl V} & 7 / 7 & 0.760 & 0.003257 \\
 & {\sl H} & 92 / 93 & 1.700 & 0.003257 \\
IRSF & {\sl J} & 3 / 3 & 1.000 & 0.000  \\
 & {\sl H} & 3 / 3 & 1.000 & 0.000  \\
 & {\sl $K_{S}$} & 3 / 3 & 1.000 & 0.000 \\
{\it Spitzer} & $L$ & 21 / 21 & 3.120 & 0.003257 \\
\enddata
\end{deluxetable}

\subsection{Space Based Observation}
OGLE-2017-BLG-0406 was observed by the {\it Spitzer} space telescope with the 3.6 $\mu$m ($L$-band) channel  of the IRAC camera. {\it Spitzer} started to observe this event on June 26 (HJD$^\prime$ = 7931), which was about 3 weeks after the peak because this was the first date that the target was visible in the {\it Spitzer} image. 
This event was chosen for {\it Spitzer} observations 
as part of a long-term (2014--2019) program, according to the
protocols of \citet{yee15}.  Specifically, it met objective criteria
defined by \citet{yee15}, which meant that it had to be chosen for observations
and observed at a specified cadence, independent of whether it had
a planet or not.  Accordingly, it was observed approximately once per day
for the first four weeks, but not the final two weeks of the program in 2017. 
\\
\par
In 2019, i.e., the final {\it Spitzer} microlensing season, essentially
all planetary events from 2014--2018 were observed for about a week
at baseline, primarily to check for systematics in the light curves,
in part because of concerns raised by \citet{kb19}.  See \citet{kb180029}
for further discussion.  OGLE-2017-BLG-0406 was observed seven times
under this program, meaning that there are a total of 28 data points.
As discussed in Section~\ref{sec:spitzsamp}, these seven points must be excluded
when determining whether OGLE-2017-BLG-0406Lb can enter the 
{\it Spitzer} statistical sample.  For the role of these data in the analysis
of systematic effects, see Appendix~\ref{sec:append}.
\\
\par

\section{DATA REDUCTION}
\label{sec:data}
The great majority of the ground-based data were reduced using the pipelines developed by the individual collaborations based on difference image analysis (DIA) method developed by \citet{tom96,ala98}. The OGLE {\sl I}-band data were reduced by the OGLE DIA \citep{woz00} photometry pipeline \citep{uda15b}. The MOA-Red and {\sl V-}band data were reduced by the MOA DIA pipeline \citep{bon01}. KMTNet-{\sl I} band data were reduced with their pySIS photometry pipeline \citep{alb09}. $\mu$FUN data were reduced using DoPhot \citep{sch93}, and LCO data were reduced using pySIS \citep{alb09}. Danish data were reduced using an updated version of DanDIA \citep{bra08}. IRSF images were reduced using the standard IRSF pipeline and MOA DIA pipeline. $\it Spitzer$ $L$-band data was reduced using methods described in \citet{cal15}. 
\\
\par
The error bars must be renormalized to accurately estimate the uncertainties. We use the following formula to rescale the errors, $\sigma'_{i} = k \sqrt{\sigma^2_{i} + e^2_{\sl min}}$, where $\sigma_{i}$ and $\sigma'_{i}$ are original and renormalized error bars in magnitudes, and $k$ and $e_{min}$ are rescaling factors \citep{ben08}. The value of $e_{min}$ represents systematic errors that dominate at high magnification or when the target is very bright. First, we fit all the light curves to find a tentative best-fit model. Then we apply $e_{min}=0.003257$ and choose $k$ values to give $\chi^2/dof = 1$ for a preliminary best-fit model. Finally, all the normalized light curves are fit again and we get the final best-model. In this process, we find that there are systematics in the data from Danish, Craigie, Farm Cove and PEST. Also Danish data are not consistent with the OGLE and LCO-CTIO data. Hence, they are not used for the analysis. We note that the data from KMT-S are not used because of the systematics that mimic the parallax signal as described in Section~\ref{sec:anal}. The data sets we used, together with the values of $k$ and $e_{\sl min}$, are shown in Table \ref{tab1}. We also note that for the OGLE-IV data, we check the standard error correction procedure described in \citet{sko16} and find that the resulting error bars are similar to those estimated in this work. We choose $k=1$ and $e_{\rm min}=0$ for IRSF data.
\\
\par

\section{GROUND-BASED LIGHT CURVE ANALYSIS}
\label{sec:anal}
For the point-source point-lens (PSPL) model, one needs three parameters to characterize the microlens light curve: $t_{0}$, the time of closest approach of the source to the lens mass; $u_{0}$, the impact parameter in units of the angular Einstein radius $\theta_{\rm E}$; and $t_{\rm E}$, the Einstein radius crossing time. For the binary-lens model, one needs three additional parameters: $q$, the planet / host mass ratio; $s$, the projected planet - star separation in units of the Einstein radius; and $\alpha$, the angle of the source trajectory relative to the binary lens axis. When we take account of the finite source effect and the parallax effect, the angular radius of the source star in units of $\theta_{\rm E}$, $\rho$, and the north and east components of the microlensing parallax vector, $\pi_{\rm E,N}$ and $\pi_{\rm E,E}$, are added for each case. The model light curve is given by
\begin{equation}
F(t) = A(t)F_{{\rm S}, i} + F_{{\rm b}, i},
\end{equation}
where $F(t)$ is the flux at time $t$, $A(t)$ is the magnification of the source star at $t$, and $F_{{\rm S}, i}$ and $F_{{\rm b}, i}$ are the baseline fluxes from the source and blend stars for each data set, $i$, respectively.
\\
\par
We use linear limb darkening models for the source star. The effective temperature of the source star estimated from the extinction corrected source color, $(V - I)_{\rm S,0}=1.02$ as discribed in Section~\ref{sec:cmd}, is $T_{\rm eff} \sim 4848 $K \citep{gon09}. Rounding to the nearest $T_{\rm eff}$ given in \citet{cla00} and assuming surface gravity $\log [g / (\rm cm \ s^{-2})] = 4.5$, and metallicity $\log [\rm M/H] = 0$, we selected limb darkening coefficients $u_{\lambda}$ to be $u_{I} = 0.6049$, $u_{Red} = 0.6534$, $u_{V} = 0.7796$, $u_{R} = 0.7081$, $u_{J} = 0.4896$, $u_{H} = 0.4252$ and $u_{K_{S}} = 0.3642$, respectively \citep{cla00}. The MOA-Red value is the mean of the {\sl R-} and {\sl I-}band values.
\\
\par
We first conduct the light curve modeling by only using ground-based data. Our light curve modeling was done using the image-centered ray-shooting method \citep{ben96,ben10} and the Markov Chain Monte Carlo (MCMC) algorithm \citep{ver03}. Note that the source and blend flux parameters are not MCMC parameters, but are fit linearly to each model following \citet{rhi99}. To find the global best-fit model, we first conduct a grid search by fixing three parameters  ($q$, $s$, $\alpha$) while the other parameters ($t_0$, $t_{\rm E}$, $u_0$, $\rho$) allowed to be free.  Next, we search for the best-fit model by refining all parameters for those models with the 100 smallest $\chi^2$ values as initial parameters. From this modeling, we find a planetary model that  has the best-fit values of $q \sim 0.0007$ and $s \sim 1.128$. 
The best-fit parameters are shown in Table \ref{tab:standard}. Because the source crosses the central caustic very close to the lens host star as seen in Figure \ref{fig:lc}, the finite source effect is well measured. The $\Delta \chi^2$ between the best-fit model and the single lens model is more than 20000. Thus, the planetary signal is detected confidently. We also explore the binary-source single-lens model \citep{gau98} and find that the model is ruled out by $\Delta \chi^2 > 3000 $.
\\
\par

\begin{deluxetable}{lcc}
\tablecolumns{3} \tablewidth{0pc} \tablecaption{\textsc{Standard
models}} \tablehead{ \colhead{Parameters} & \colhead{Wide~($s>1$)} &
\colhead{Close~($s<1$)} } \startdata
  $\chi^2/\rm{dof}$               &29272.22/29285       &29325.836/29285         \\
  $t_0$ $(\rm{HJD}^{\prime})$     &7908.809 $\pm$ 0.001  &7908.810 $\pm$ 0.001  \\
  $u_0$ $(10^{-3})$               &9.437 $\pm$ 0.028     &9.454 $\pm$ 0.029    \\
  $t_{\rm E}$ $(\rm{days})$       &37.043 $\pm$ 0.082    &36.994 $\pm$ 0.083    \\
  $s$                             &1.129 $\pm$ 0.001     &0.895 $\pm$ 0.001     \\
  $q$ $(10^{-4})$                 &7.024 $\pm$ 0.090     &6.876 $\pm$ 0.091     \\
  $\alpha$ $(\rm{rad})$           &0.993 $\pm$ 0.001     &0.993 $\pm$ 0.002    \\
  $\rho$ $(10^{-3})$              &5.861 $\pm$ 0.025     &5.830 $\pm$ 0.025     \\
  $f_S({\rm OGLE})$               &1.462 $\pm$ 0.004     &1.464 $\pm$ 0.004     \\
  $f_B({\rm OGLE})$               &0.102 $\pm$ 0.004     &0.100 $\pm$ 0.004     \\
  $t_*$ $(\rm{days})$             &0.217 $\pm$ 0.001     &0.216 $\pm$ 0.001     \\
\enddata
\tablecomments{$t_*\equiv\rho t_\e$ is a derived quantity and is
not fitted independently. All fluxes are on an 18th magnitude scale,
e.g., $I_s= 18-2.5\,\log(f_s)$.}
\label{tab:standard}
\end{deluxetable}

High magnification planetary microlensing  events often have a so-called ``close-wide" degeneracy because the structures near the central caustic are very similar to each other for $s \leftrightarrow s^{-1}$, particularly for $s \ll 1$ and $s \gg 1$ \citep{gri98, dom99, chu05}. We search for the model with $s < 1$ and find the model that has the best-fit values of $q \sim 0.0007$ and $s \sim 0.895$. But this close model has worse $\chi^2$ compared to the best-fit wide model by $\Delta \chi^2 \sim 41$. This difference mostly comes from the data near the peak. Thus, we exclude the close model because of its poor fit. 
\\
\par
\begin{figure}
\plotone{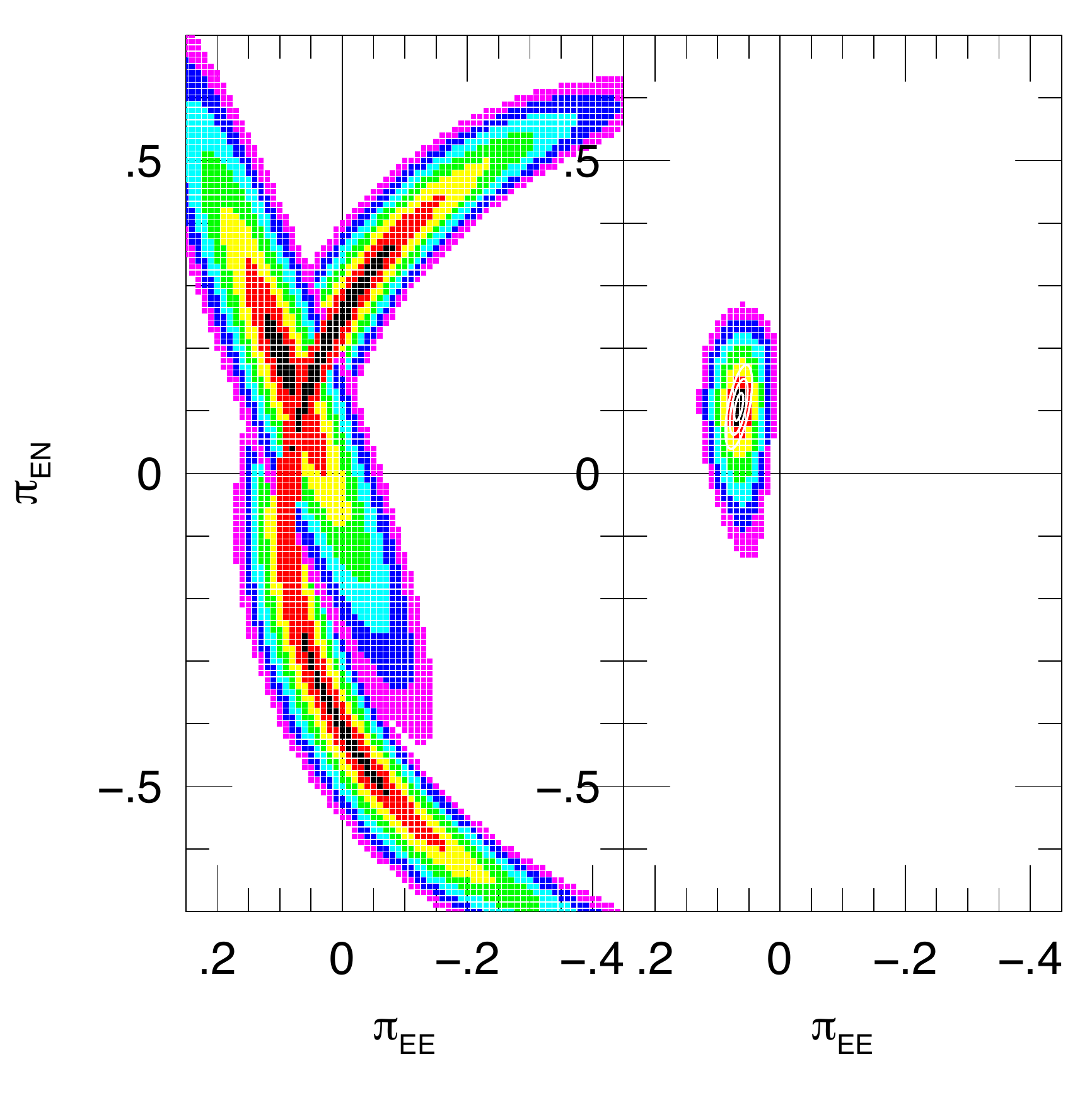}
\caption{OGLE-2017-BLG-0406 parallax contours for the W+ solution. 
Left: the ground-only 
(elliptical) contours are derived from the covariance matrix from the
MCMC, while the {\it Spitzer}-``only'' (arc-like) contours are derived from
the analytical expression in Equation~(\ref{eqn:linear3}).  The colors
(black, red, yellow, green, cyan, blue, magenta) represent
$\Delta\chi^2<(1,4,9,16,25,36,49)$.  Note that the $1\,\sigma$ contours
for the ground-only and {\it Spitzer}-``only'' measurements overlap.
Right: Colored contours are the $\Delta\chi^2$ values for the sum of
the two $\chi^2$ distributions that are shown to the left.  Despite
the fact the each set of contours on the left provides essentially
1-D information, the combination is well constrained in both dimensions.
The white ellipses represent the $1\,\sigma$, $2\,\sigma$, and $3\,\sigma$
contours from the combined numerical fit to all of the data.  The
semi-analytic approach (colored contours) provides a very good, although
not perfect, representation of the full numerical result.  This
shows that the semi-analytic approach enables one to accurately
trace the information flow.
}
\label{fig:arc+}
\end{figure}
\begin{figure}
  \plotone{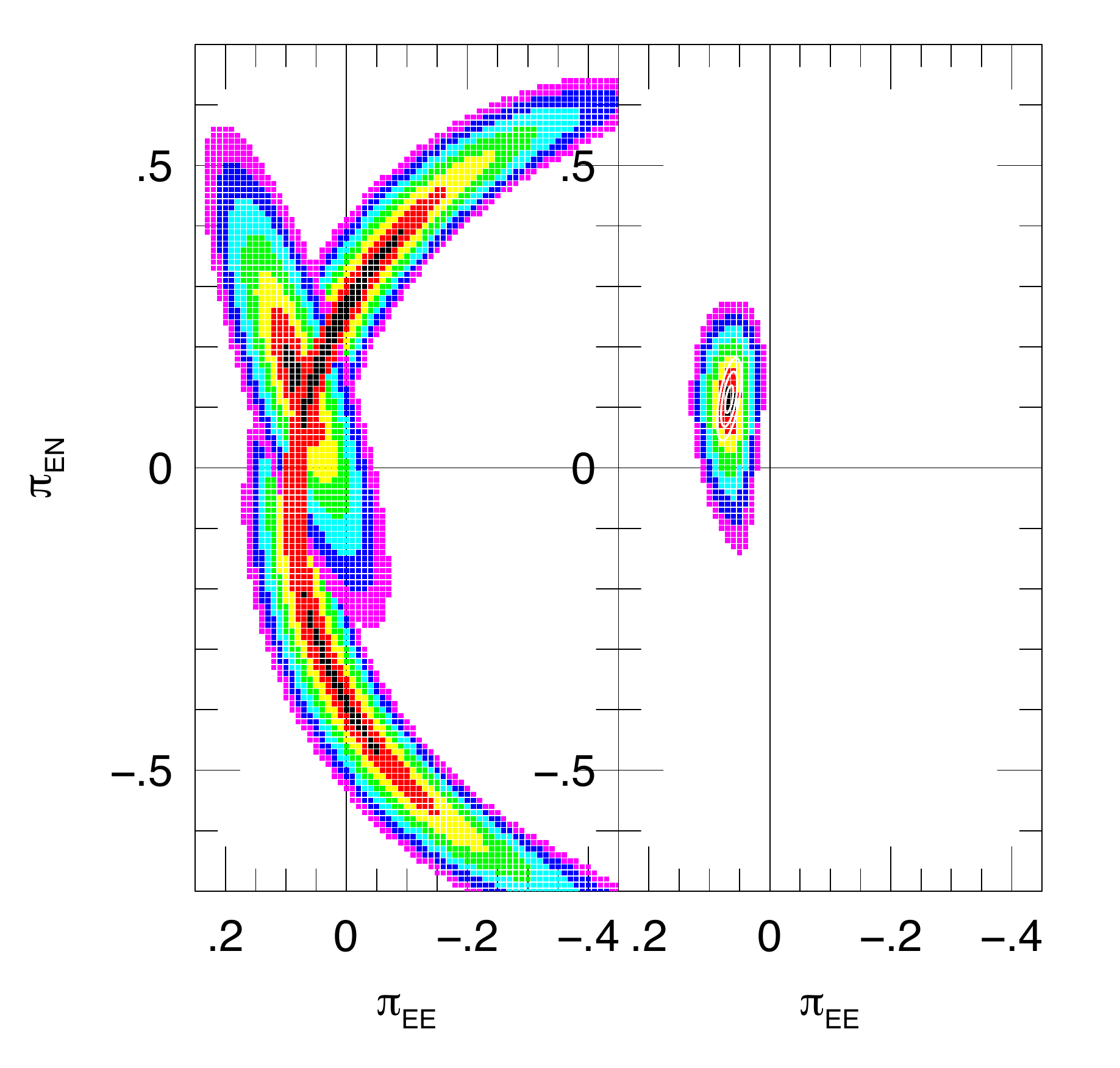}
\caption{OGLE-2017-BLG-0406 parallax contours for the W- solution.
Similar to Figure~\ref{fig:arc+}.
}
\label{fig:arc-}
\end{figure}

When $t_{\rm E}$ is large, we have a chance to measure the orbital parallax effect, which is caused by the acceleration of Earth's orbital motion \citep{gould92,alc95}. We do not expect a significant orbital microlensing parallax signal for such a short event, in the middle of the season because the acceleration of Earth projected to the bulge is the smallest. 
We begin by doing a parallax fit without the {\it Spitzer} data to independently assess parallax constraints coming from the ground-based data.
We conduct the parallax fit by adding the two additional parameters, $\pi_{\rm E,N}$ and $\pi_{\rm E,E}$. In the first iteration, we found a model with a large $\pi_{\rm E}$ value of $\sim 0.4$. However, the $\Delta \chi^2$ between the standard model and the parallax model comes mostly from KMT-S data, and it was not consistent with the other data sets. Also, the differences were from the baseline. Hence, we conduct parallax analysis without the KMT-S data set because we think that there is a systematic error in the data set, which mimics the deviation caused by the parallax effect. Then we tried the parallax fit again and obtained a smaller $\pi_{\rm E}$ value of $\sim$ 0.2. The best-fit parameters are shown in Table \ref{tab:parallax}. 
While the improvement in $\chi^2$ is relatively small ($\Delta\chi^2=6.9$),
Figures~\ref{fig:arc+} and \ref{fig:arc-} show that there is a strong
one-dimensional (1-D) parallax constraint, which arises from the asymmetry
in the light curve induced by the instantaneous acceleration of
Earth around the peak of the event \citep{gmb94}.  The relatively
small $\Delta\chi^2$ simply reflects the fact that this 1-D constraint
happend to pass close to the origin.  We will return to the role of this
1-D parallax constraint after including {\it Spitzer} data into the analysis.
\\
\par

\begin{deluxetable}{lcccc}
\tablecolumns{5} \tablewidth{0pc}\rotate
\tablecaption{\textsc{Parallax models for ground-only data}}
\tablehead{ \colhead{Parameters } & \colhead{Wide$(+)$} &
\colhead{Wide$(-)$} &\colhead{Close$(+)$} &\colhead{Close$(-)$}}
\startdata
  $\chi^2/\rm{dof}$            &29272.225/29283           &29273.381/29283           &29318.530/29283           &29318.828/29283\\
  $t_0$ $(\rm{HJD}^{\prime})$  &7908.809 $\pm$ 0.001      &7908.809 $\pm$ 0.001      &7908.809 $\pm$ 0.001      &7908.809 $\pm$ 0.001\\
  $u_0$ $(10^{-3})$            &9.402 $\pm$ 0.032         &-9.447 $\pm$ 0.030        &9.432 $\pm$ 0.027         &-9.419 $\pm$ 0.030\\
  $t_{\rm E}$ $(\rm{days})$    &37.168 $\pm$ 0.096        &37.042 $\pm$ 0.091        &37.075 $\pm$ 0.076        &37.119 $\pm$ 0.091\\
  $s$                          &1.129 $\pm$ 0.001         &1.130 $\pm$ 0.001         &0.895 $\pm$ 0.001         &0.895 $\pm$ 0.001\\
  $q$ $(10^{-4})$              &6.972 $\pm$ 0.093         &7.073 $\pm$ 0.092         &6.894 $\pm$ 0.089         &6.831 $\pm$ 0.090\\
  $\alpha$ $(\rm{rad})$        &0.991 $\pm$ 0.001         &-0.992 $\pm$ 0.002        &0.993 $\pm$ 0.002         &-0.992 $\pm$ 0.002\\
  $\rho$ $(10^{-3})$           &5.832 $\pm$ 0.025         &5.874 $\pm$ 0.025         &5.822 $\pm$ 0.025         &5.809 $\pm$ 0.025\\
  $\pi_{\rm{E},\it{N}}$        &0.179(0.167) $\pm$ 0.094  &0.157(0.168) $\pm$ 0.066  &0.137(0.121) $\pm$ 0.031  &0.198(0.168) $\pm$ 0.065\\
  $\pi_{\rm{E},\it{E}}$        &0.097(0.089) $\pm$ 0.037  &0.097(0.088) $\pm$ 0.026  &0.064(0.070) $\pm$ 0.014  &0.098(0.086) $\pm$ 0.027\\
  $\pi_{\rm{E}}$               &0.204(0.193) $\pm$ 0.093  &0.185(0.191) $\pm$ 0.069  &0.151(0.141) $\pm$ 0.030  &0.221(0.189) $\pm$ 0.068\\
  $\phi_\pi$                   &0.494(0.554) $\pm$ 0.481  &0.554(0.507) $\pm$ 0.123  &0.438(0.538) $\pm$ 0.124  &0.459(0.498) $\pm$ 0.155\\
  $f_S({\rm CTIO})$            &1.457 $\pm$ 0.004         &1.463 $\pm$ 0.004         &1.460 $\pm$ 0.004         &1.459 $\pm$ 0.004\\
  $f_B({\rm CTIO})$            &0.107 $\pm$ 0.004         &0.101 $\pm$ 0.004         &0.104 $\pm$ 0.003         &0.105 $\pm$ 0.004\\
  $t_*$ $(\rm{days})$          &0.217 $\pm$ 0.001         &0.218 $\pm$ 0.001         &0.216 $\pm$ 0.001         &0.216 $\pm$ 0.001\\
\enddata
\tablecomments{Mean values from the MCMC are shown in parentheses.
All other values are from the best-fit model.
$\pi_\e\equiv\sqrt{\pi_{\e,N}^2 + \pi_{\e,E}^2}$,
$\phi_\pi\equiv\tan^{-1}(\pi_{\e,E}/\pi_{\e,N})$, and $t_*\equiv\rho
t_\e$ are derived quantities and are not fitted independently.  All
fluxes are on an 18th magnitude scale, e.g., $I_s=
18-2.5\,\log(f_s)$.} \label{tab:parallax}
\end{deluxetable}

\section{{{\it Spitzer} Parallax Analysis}
\label{sec:spitzpar}}

\subsection{{{\it Spitzer}-``only'' Parallax}
\label{sec:spitz-only}}
In principle, we could now proceed to incorporate the {\it Spitzer}
data into a joint fit together with the ground-based data.  We will
do so in Section~\ref{sec:spitz-joint}.  
First, however, it is important to 
examine how the {\it Spitzer} data and the ground-based
data contribute information to the parallax measurement.  The principal
reason for doing so is that both data sets can be subject to systematic
errors, which are of very different types and can affect the parallax
measurement very differently.  An important check for such systematics
is whether the parallax information derived from each data set is consistent
with the other.
Failure of this test would provide clear evidence for systematics in
one or both data sets.  In addition, we will ultimately be making
somewhat separate use of the magnitude and direction of the parallax
vector.  In order to understand how secure each of these components is,
we will need to trace their origins in different combinations of
{\it Spitzer}-based and ground-based information.
\\
\par
In Section~\ref{sec:anal}, we showed that the ground-based data yielded
essentially one-dimensional parallax information, with $\pi_{\e,\parallel}$
(the component parallel to the instantaneous projected direction of the
Sun at $t_0$) measured about nine (for W+) or five (for W-) 
times more precisely than the
orthogonal component $\pi_{\e,\perp}$.  This is because the information
for the latter comes from further out in the wings of the light curve
\citep{smp03,gould04}, which, for events like OGLE-2017-BLG-0406 that are not
extremely long, is generally quite faint.  This also means that the
$\pi_{\e,\perp}$ component is much more sensitive to long-term trends in the data.
In the present case, for which $\pi_{\e,\parallel}\simeq 0$, this means
that the direction of the ground-based parallax vector is determined
much more confidently than its magnitude.
\\
\par
The {\it Spitzer} light curves can also be affected by long term trends
in the data, which can affect the parallax measurement.  In their analysis
of 50 {\it Spitzer} events from 2015, \citet{zhu17} identified five
with obvious trends in the data, and \citet{kb19} identified 14 more.
There is only one case for which the causes of such trends have been
investigated: KMT-2018-BLG-0029 \citep{kb180029}.  In that case,
bright nearby blends with poorly determined positions were found
to be likely to have generated trends as the {\it Spitzer} camera 
rotated during the season.
Because the source was faint and not well magnified, the trends were
about 30\% of the total observed flux variations.
Nevertheless, after the trends were removed, the amplitude of the parallax
measurement only changed by 20\% (about $2\,\sigma$).  Thus, it is important
to both carefully evaluate and minimize the impact of potential
{\it Spitzer} systematics.
\\
\par
\citet{refsdal66} originally analyzed satellite parallaxes prior to
the time \citep{gould92} that it was recognized that the ground-based
light curve alone would have any parallax information.  Hence, although
not explicitly stated, his was in essence a satellite-``only'' analysis.
The ground-based parameters $(t_{0,\oplus},u_{0,\oplus},t_{\e,\oplus})$ were
directly compared to the satellite parameters 
$(t_{0,\sat},u_{0,\sat},t_{\e,\sat})$ to produce the parallax measurement
\begin{equation}
\bpi_\e = {\au \over D_\perp}(\Delta\tau,\Delta u_0);
\qquad
\Delta\tau \equiv {t_{0,\sat} - t_{0,\oplus}\over t_\e};
\qquad
\Delta u_0 \equiv {u_{0,\sat} - u_{0,\oplus}},
\label{eqn:refsdal}
\end{equation}
where ${\bf D}_\perp$ is the projected separation of the satellite from Earth
and where it was implicitly assumed that the Einstein timescales were the
same $t_\e = t_{\e,\oplus}= t_{\e,\sat}$.  The implicit idea (as illustrated
in Figure~1 of \citealt{gould94b} and first realized in Figure~1 of
\citealt{ob140939}) is that $(t_{0,\sat},u_{0,\sat},t_{\e,\sat})$ would
be ``observables'' from the satellite, just as the corresponding quantities
were from Earth.  Note that Equation~(\ref{eqn:refsdal}) has a four-fold
degeneracy because $u_0$ is a signed quantity but only $|u_0|$ is
determined directly from the light curve.  See Figure~4 of \citet{gould04}
for the sign convention.
\\
\par
However, in real {\it Spitzer} microlensing events, the peak is very
often not observed from space, primarily because there is a $\sim$ 3--10 day time 
delay between identifying the event and initiating satellite
observations (Figure~1 from \citealt{ob140124}).  Hence, while
Equation~(\ref{eqn:refsdal}) remains formally valid, it may no longer
express the parallax measurement in terms of ``observables'',
because $(t_0,u_0)_\sat$ may not be separately determined.
\\
\par
\citet{gould19} generalized Refsdal's satellite-``only'' analysis
to the case of satellite data streams that did not cover the peak.
He showed that if the source flux in the satellite observations
was known and the baseline flux was measured, then each satellite measurement
at finite magnification yields an exactly circular degeneracy in the
$\bpi_\e$ plane.  In the presence of measurement errors these circles
become finite annuli.  If these measurements cover the peak, then
the corresponding circles intersect in exactly two places, which then 
reproduces the \citet{refsdal66} four-fold degeneracy (two pairs, one each for
$\pm u_{0,\oplus}$).  See Figure~3 of \citet{gould19}.  On the other hand,
the osculating circles from a series of late-time measurements combine
to form an extended arc.  See Figure~1 of \citet{gould19}.  Such
arcs may yield exquisite 1-D constraints on $\bpi_\e$ while still
providing almost no constraint on its amplitude $\pi_\e$.  See second row
of Figure~2 from \citet{kojima2} for an extreme example.  Nevertheless,
as that example makes clear, the addition of exterior information about the 
direction of $\bpi_\e$ can then constrain $\pi_\e$ very well.
The first case of such an arc appearing in a {\it Spitzer}-``only'' analysis
was OGLE-2018-BLG-0596 \citep{ob180596}.  OGLE-2017-BLG-0406 has a 
qualitatively similar arc-like degeneracy.
\\
\par
We note that if $(t_0,u_0,t_\e)_\oplus$ are considered as known exactly,
then for each trial value of $\bpi_\e=(\pi_{\e,N},\pi_{\e,E})$, the remaining
two parameters $(f_s,f_b)_{Spitzer}$ can be calculated analytically.
That is, there are $(N_{Spitzer}+1)$ linear equations for two unknowns,
where $N_{Spitzer}$ is the number of {\it Spitzer} measurements, $F(t_k)$.
These are $N_{Spitzer}$ equations for the measurements,
\begin{equation}
y_k = \sum_{i=1}^2 a_i g_{i,k} \pm \sigma_k
\qquad
y_k\equiv F(t_k),
\quad
a\equiv (f_s,f_b)_{Spitzer};
\quad
g_{1,k} \equiv A(t_k),
\quad
g_{2,k} \equiv 1.
\label{eqn:linear}
\end{equation}
plus one for the flux constraint
\begin{equation}
y_0 = \sum_{i=1}^2 a_i g_{i,0} \pm \sigma_0
\qquad
y_0\equiv f_{s,spitzer,{\rm constr}},
\quad
g_{1,0} \equiv 1
\quad
g_{2,k} \equiv 0.
\label{eqn:linear2}.
\end{equation}
Then one solves in the usual way,
\begin{equation}
d_i = \sum_{\mu=0}^{N_{Spitzer}} {y_\mu g_{i,\mu}\over \sigma_\mu^2};
\qquad
b_{i,j} = \sum_{\mu=0}^{N_{Spitzer}} {g_{i,\mu}g_{j,\mu}\over \sigma_\mu^2};
\qquad
c = b^{-1}
\qquad
a_i = \sum_j c_{i,j} d_j,
\label{eqn:linear3}
\end{equation}
with $c_{i,j}$ being the covariance matrix of the two parameters.
\\
\par
To evaluate the {\it Spitzer}-``only'' parallax contours, we 
calculate $A_{Spitzer}(t_k)$ by fixing $(t_0,u_0,t_\e)_\oplus$ according
to the W+ and W- solutions shown in Table~\ref{tab:parallax} 
and by fixing $\bpi_\e = (\pi_{\e,N},\pi_{\e,E})$ at a grid of values.
In Section~\ref{sec:cmd}, we evaluate $f_{s,spitzer,{\rm constr}} = 11.10 \pm 0.15$.
We find that the {\it Spitzer} errors must be renormalized by a factor
3.4 to achieve $\chi^2/{\rm dof} = 1$.
The arc in Figure~\ref{fig:arc+} shows the resulting {\it Spitzer}-``only'' 
contours for the W+ solution. The diagonal contours represent
the ground-only parallax measurement, which we have extended out
to seven sigma analytically using the covariance matrix from the MCMC.
Figure~\ref{fig:arc-} shows the corresponding structures for the W-
solution.
\\
\par
The most important feature is that the $1\,\sigma$ contours from the
two parallax measurements overlap.  Hence, there is no tension
at all between the two determinations.  Second, the direction of
$\bpi_\e$ is essentially determined by the ground-based measurement.
That is, even if the arc were displaced to the East or West, it
would intersect the ground contours at a very similar polar angle.
The only exception would be if it intersected very close to the
origin.  Third, the best-fit ground-based value of $\pi_{\e,\perp}$ plays
very little role in the point of overlap of these two sets of contours,
which is shown in the right hand panel of the figure.  That is,
even if $\pi_{\e,\perp}$ were displaced two sigma toward higher values,
the overlap would occur in the same place.  This means that the
aspect of the ground-based data that is most vulnerable to systematic
errors does not play much role in the final solution.  The panel
at the right shows that despite the fact that the ground-only and
{\it Spitzer}-``only'' measurements are effectively 1-D, they
combine to form tight 2-D constraints.
\\
\par

\subsection{{Combined {\it Spitzer} and Ground-based Analysis}
\label{sec:spitz-joint}}
We therefore proceed to directly analyze the ground-based and {\it Spitzer}
data jointly.  The resulting microlensing-parameter estimates are
given in Table~\ref{tab:combined}, where in particular, we show two different
representations of the parallax vector $\bpi_\e$, i.e., in Cartesian
$(\pi_{\e,N},\pi_{\e,E})$ and polar $(\pi_\e,\pi_\phi)$ coordinates.
We evaluate the $\bpi_\e$ covariance matrix and use this to generate
$1\,\sigma$, $2\,\sigma$, and $3\,\sigma$ contours, which are
shown in the right panels Figures~\ref{fig:arc+} and \ref{fig:arc-}
as white ellipses.  These show that the semi-analytic approach
described in Section~\ref{sec:spitz-only} and displayed in these figures
works quite well, although not perfectly.  This good agreement confirms
that there is strong physical basis for the arguments given in that
section.
\\
\par

\begin{deluxetable}{lcc}
\tablecolumns{3} \tablewidth{0pc} \tablecaption{\textsc{Wide models
for ground+{\it Spitzer} data}} \tablehead{ \colhead{Parameters } &
\colhead{Wide$(+,+)$} & \colhead{Wide$(-,+)$} } \startdata
  $\chi^2/\rm{dof}$               &29297.755/29310          &29295.132/29310       \\
  $t_0$ $(\rm{HJD}^{\prime})$     &7908.813 $\pm$ 0.001     &7908.813 $\pm$ 0.001  \\
  $u_0$ $(10^{-3})$               &9.281 $\pm$ 0.028        &-9.281 $\pm$ 0.028    \\
  $t_{\rm E}$ $(\rm{days})$       &37.134 $\pm$ 0.083       &37.133 $\pm$ 0.085    \\
  $s$                             &1.128 $\pm$ 0.001        &1.128 $\pm$ 0.001     \\
  $q$ $(10^{-4})$                 &6.955 $\pm$ 0.061        &6.970 $\pm$ 0.090     \\
  $\alpha$ $(\rm{rad})$           &0.993 $\pm$ 0.001        &-0.993 $\pm$ 0.001    \\
  $\rho$ $(10^{-3})$              &5.852 $\pm$ 0.025        &5.843 $\pm$ 0.025     \\
  $\pi_{\rm{E},\it{N}}$           &0.126(0.111) $\pm$ 0.021 &0.120(0.113) $\pm$ 0.024  \\
  $\pi_{\rm{E},\it{E}}$           &0.062(0.066) $\pm$ 0.007 &0.065(0.067) $\pm$ 0.007  \\
  $\pi_{\rm{E}}$                  &0.140(0.130) $\pm$ 0.016 &0.136(0.133) $\pm$ 0.018  \\
  $\phi_\pi$                      &0.455(0.549) $\pm$ 0.127 &0.499(0.549) $\pm$ 0.134  \\
  $f_S({\rm OGLE})$               &1.459 $\pm$ 0.004        &1.459 $\pm$ 0.004     \\
  $f_B({\rm OGLE})$               &0.106 $\pm$ 0.004        &0.105 $\pm$ 0.004     \\
  $f_S(Spitzer)$                  &11.249 $\pm$ 0.164       &11.210 $\pm$ 0.180    \\
  $f_B(Spitzer)$                  &-2.656 $\pm$ 0.165       &-2.614 $\pm$ 0.182    \\
  $t_*$ $(\rm{days})$             &0.217 $\pm$ 0.001        &0.217 $\pm$ 0.001     \\
\enddata
\tablecomments{Mean values from the MCMC are shown in parentheses.
All other values are from the best-fit model. $\pi_\e$, $\phi_\pi$,
and $t_*$ are derived quantities and are not fitted independently.
All fluxes are on an 18th magnitude scale, e.g., $I_s=
18-2.5\,\log(f_s)$.} \label{tab:combined}
\end{deluxetable}

\section{{COLOR-Magnitude Diagram}
\label{sec:cmd}}
We can derive the angular Einstein radius, $\theta_{\rm E} = \theta_{\star} / \rho$, because the finite source size, $\rho$, is constrained from the light-curve fitting and the angular size of the source star, $\theta_{\star}$, can be derived from the extinction-corrected source color and brightness. The measurement of $\theta_{\rm E}$ gives the following mass-distance relation of lens system,
\begin{equation}
M = \frac{c^2}{4G}\theta_{\rm E}^2\frac{D_{\rm S}D_{\rm L}}{D_{\rm S} - D_{\rm L}} = {\theta_{\rm E}^2 \over \kappa \pi_{\rm rel}}.
\end{equation}
\\
\par

\subsection{Calibration}
We derive the source magnitudes in the $V$ and $I$ bands by converting the instrumental source magnitude in MOA-Red and MOA-V bands into the standard Kron-Cousin {\sl I-}band and Johnson {\sl V-}band scales using the following relations,
\begin{equation}
I_{\rm OGLE-III} - R_{\rm MOA} = (28.132 \pm 0.002) - (0.206 \pm 0.001)(V - R)_{\rm MOA},
\end{equation}
\begin{equation}
V_{\rm OGLE-III} - V_{\rm MOA} = (28.302 \pm 0.002) - (0.108 \pm 0.001)(V - R)_{\rm MOA}.
\end{equation}
From the light-curve fitting using these formulae, we obtain the source color and magnitude $(V-I)_{\rm S} = 2.581 \pm 0.016$ and $I_{\rm S} = 17.603 \pm 0.011$. We also calibrate the CTIO $H$-band magnitude to 2MASS \citet{car01} scale with the following relation,
\begin{equation}
H_{\rm 2mass} = H_{\rm CTIO} - 3.917 \pm 0.009
\end{equation}
based on the stars within $120^{\prime\prime}$ of the target. We find the source magnitude $H_{\rm S} = 14.696 \pm 0.010$ and derive color $(V-H)_{\rm S} = 5.488 \pm 0.016$ and $(I-H)_{\rm S} = 2.907 \pm 0.015$. 
\\
\par

\subsection{Source Angular Radius}
\begin{figure}
\begin{center}
\includegraphics[width=10cm]{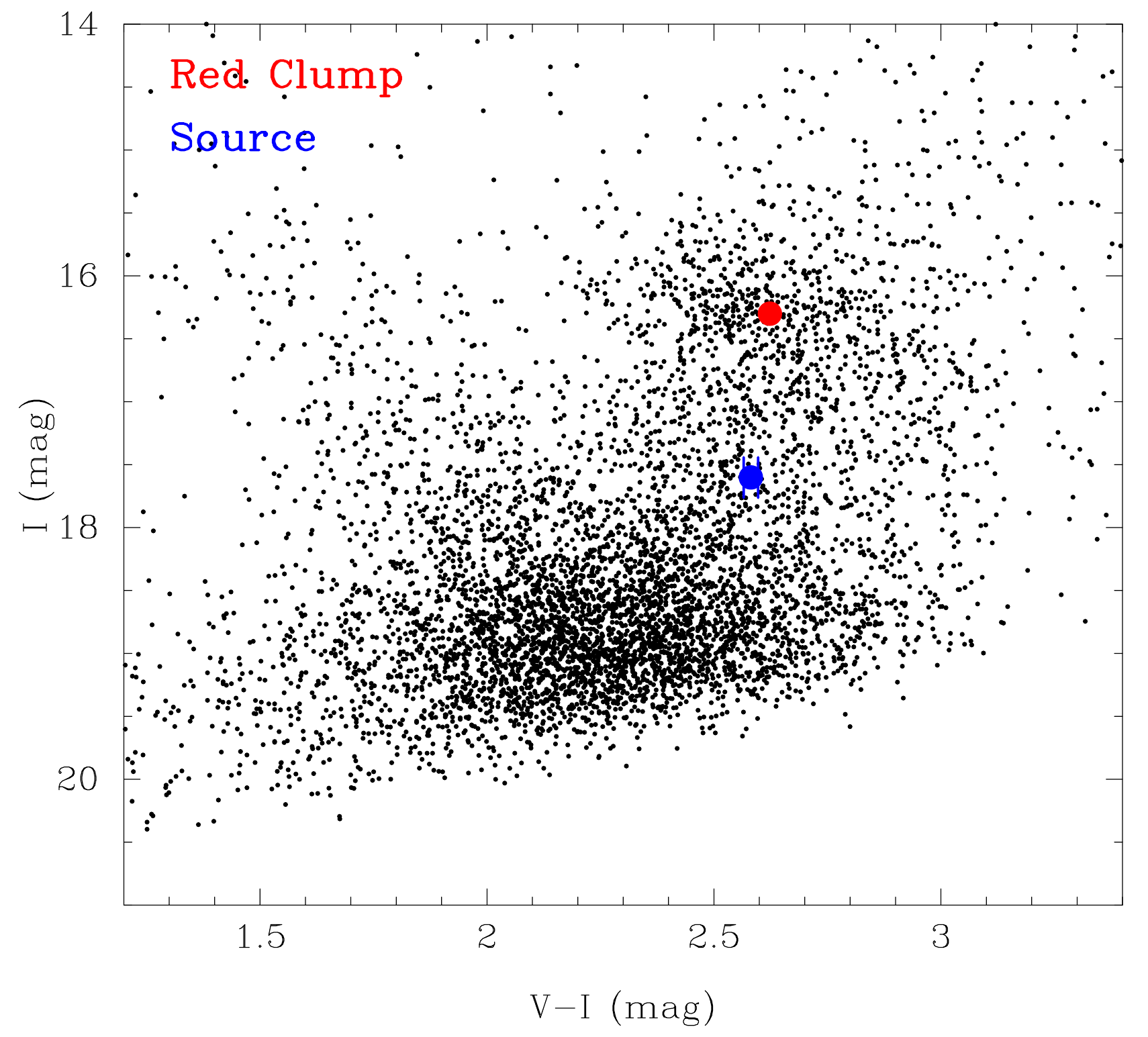}
\caption{The $(V - I , I )$ color magnitude diagram (CMD) of the OGLE stars within $120^{\prime\prime}$ of OGLE-2017-BLG-0406. The red filled circle indicates the red clump giant (RCG) centroid, and blue filled circle indicates the source color and magnitude, respectively.  \label{fig:cmd-VI}}
\end{center}
\end{figure}

\begin{figure}
\begin{center}
\includegraphics[width=10cm]{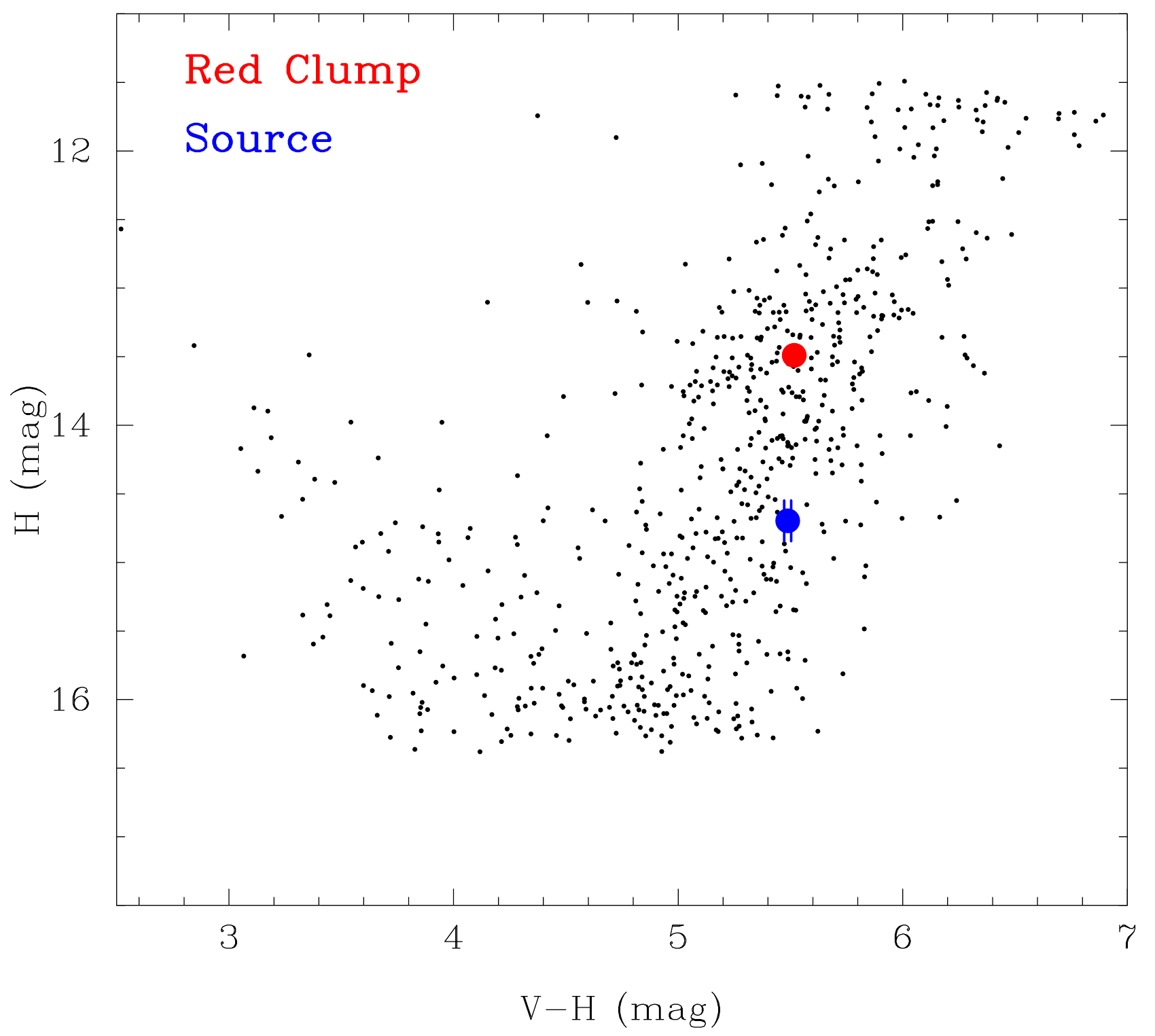}
\caption{The $(V - H , H)$ color magnitude diagram (CMD) of OGLE-2017-BLG-0406. $V$- and $H$-band magnitudes are calibrated to the Johnson $V$ and 2MASS scale, respectively. The red filled circle indicates the red clump giant (RCG) centroid, and the blue filled circle indicates the source color and magnitude, respectively.  \label{fig:cmd-VH}}
\end{center}
\end{figure}

\begin{figure}
\begin{center}
\includegraphics[width=10cm]{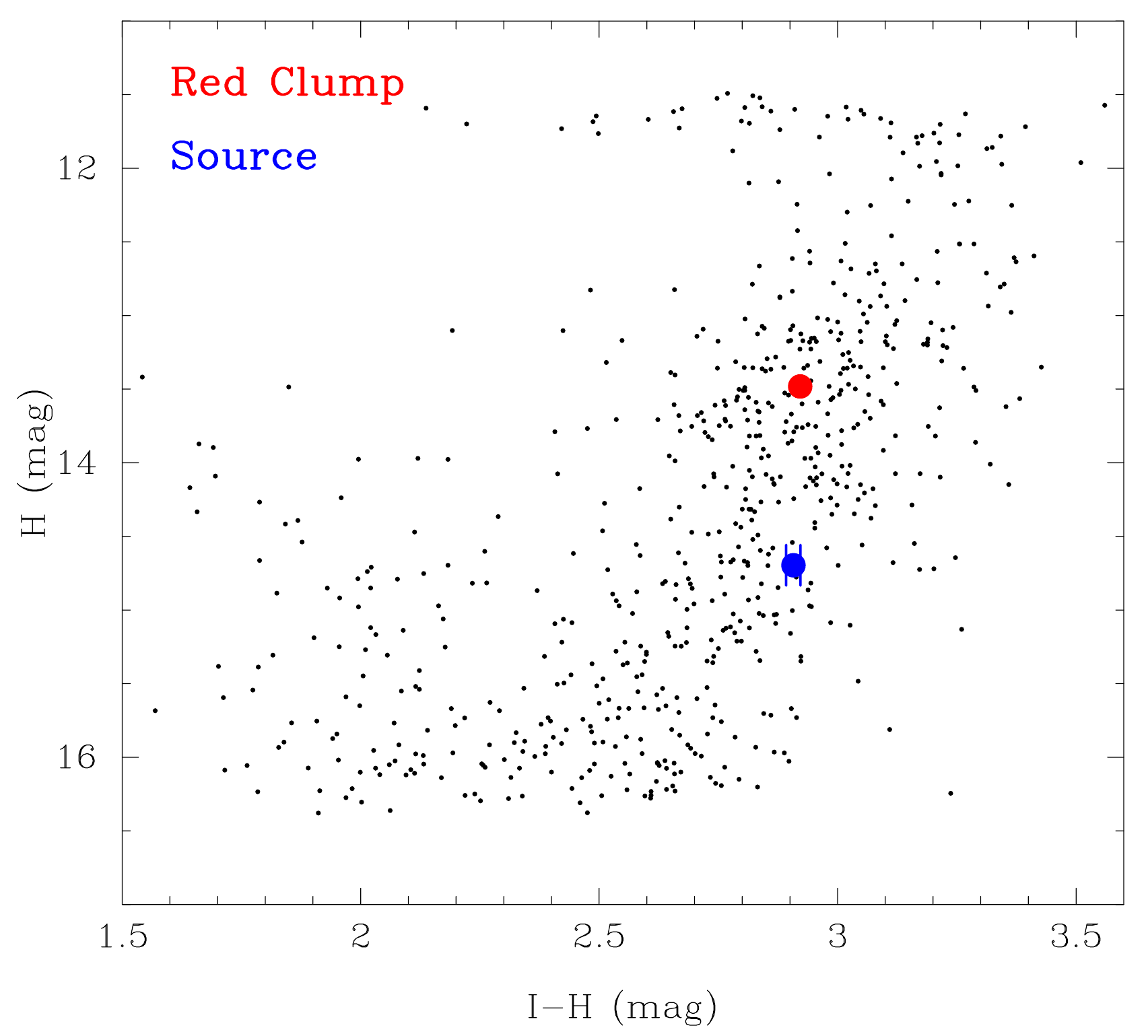}
\caption{The $(I - H , H)$ color magnitude diagram (CMD) of OGLE-2017-BLG-0406. $I$- and $H$-band magnitudes are calibrated to the Cousins $I$ and 2MASS scale, respectively. The red filled circle indicates the red clump giant (RCG) centroid, and the blue filled circle indicates the source color and magnitude, respectively.  \label{fig:cmd-IH}}
\end{center}
\end{figure}

To obtain the intrinsic source color and magnitude, we use the red clump giants (RCG) centroid in the color-magnitude diagram (CMD) as a standard candle. Figures \ref{fig:cmd-VI} to \ref{fig:cmd-IH} show CMDs of stars within $120^{\prime\prime}$ of the target. The $V$ and $I$ magnitudes are from OGLE-III catalog, and the $H$ magnitude is from the VVV catalog, which is calibrated to the 2MASS scale, respectively. We find that the centroids of the RCGs in this field, which are indicated as filled red circles, are at $I_{\rm RCG} = 16.302 \pm 0.045$, $(V-I)_{\rm RCG} = 2.623 \pm 0.012$, $(V-H)_{\rm RCG} = 5.517 \pm 0.030$ and $(I-H)_{\rm RCG} = 2.886 \pm 0.016$ from these CMDs. From \citet{nat16} and \citet{ben13}, we also find that the intrinsic magnitude and color of RCG should be $I_{\rm RCG,0} = 14.426 \pm 0.040$, $(V-I)_{\rm RCG,0} = 1.060 \pm 0.060$, $(V-H)_{\rm RCG,0} = 2.360 \pm 0.090$ and $ (I-H)_{\rm RCG,0} = 1.300 \pm 0.060$. The color and  magnitude of source and the centroid of RCG are summarized in Table \ref{tab:sc}. By subtracting the intrinsic RGC color and magnitude from the measured RGC positions in our CMDs, we find an extinction value of $A_{I, \rm obs} = 1.876 \pm 0.060$, and color excess values of $E(V-I)_{\rm obs} = 1.563 \pm 0.061$, $E(V-H)_{\rm obs} = 3.157 \pm 0.095$, and $E(I-H)_{\rm obs} = 1.586 \pm 0.062$. 
\\
\par

\begin{deluxetable}{lcccc}
\tablecaption{The source color and magnitude\label{tab:sc}}
\tablehead{
\colhead{} & \colhead{$I$} &  \colhead{$V-I$} &  \colhead{$V-H$} &  \colhead{$I-H$}  \\ 
}
\startdata
RCG (measured from CMDs) & $16.302 \pm 0.045$ & $2.623 \pm 0.012$ & $5.517 \pm 0.030$ & $2.886 \pm 0.016$  \\
RCG (extinction-corrected) & $14.426 \pm 0.040$ & $1.060 \pm 0.060$ & $2.360 \pm 0.090$ & $1.300 \pm 0.060$  \\
Source (measured from light-curve fitting) & $17.603 \pm 0.011$ & $2.581 \pm 0.016$ & $5.488 \pm 0.016$ & $2.907 \pm 0.015$  \\
Source (extinction-corrected)\tablenotemark{a}& $15.692 \pm 0.067$ & $1.020 \pm 0.055$ & $2.373 \pm 0.076$ & $1.353 \pm 0.063$  \\
\enddata
\tablenotetext{a}{Extinction-corrected magnitudes using the \citet{nis08} extinction law from Table \ref{tab:extinction}}
\end{deluxetable}

\begin{deluxetable}{ccccc}
\tablecaption{Comparison of the extinction based on different extinction laws\label{tab:extinction}}
\tablehead{
\colhead{Extinction law} & \colhead{None} & \colhead{Cardelli et al.(1998)} & \colhead{Nishiyama et al.(2009)} & \colhead{Nishiyama et al.(2008)} \\
}
\startdata
$A_V$      & $ 3.437 \pm  0.086$ &  $3.565 \pm 0.055$   &  $3.497 \pm 0.062$    & $\bf 3.472 \pm 0.082$ \\
$A_I$        & $ 1.876 \pm 0.060$ & $ 1.982 \pm 0.050$   &  $1.931 \pm 0.050$    & $\bf 1.911 \pm 0.066$ \\
$A_H$       & $ 0.364 \pm 0.103$ &  $0.583 \pm 0.018$   &  $0.467 \pm 0.012$    & $\bf 0.358 \pm 0.009$ \\
$E(V-I)$      &  $1.563 \pm 0.061$ &  $1.587 \pm 0.037$   &  $1.566 \pm 0.038$    & $\bf 1.561 \pm 0.053$ \\
$E(V-H)$     &  $3.157 \pm 0.095$ &  $2.987 \pm 0.047$   &  $3.031 \pm 0.056$    & $\bf 3.115 \pm 0.074$ \\
$E(I-H)$      &  $1.586 \pm 0.062$ &  $1.401 \pm 0.033 $  &  $1.464 \pm 0.052$   & $\bf 1.554 \pm 0.061 $\\
$\chi^2 / {\rm dof}$ &  -               & 11.60/1             &  2.70/1             & 2.66/2          \\
\enddata
\end{deluxetable}

The extinction can be determined most accurately if three colors are used \citep{ben10b}. Following \citet{ben10b} and \citet{kos17}, we fit them with the extinction law of \citet{car98}, and \citet{nis08,nis09}. Table \ref{tab:extinction} shows the results of fitting extinction values to those of extinction laws. We adopt  $R_{JKVI} \equiv E(J-Ks)/E(V-I) = 0.3347$ value from \citet{nat13} for the event coordinates. We see that the $\chi^2$ value using \citet{nis08} extinction law is the smallest and the extinction values agree with our measurement from our CMDs. Thus, we decide to use the results from \citet{nis08} extinction law for the rest of the analysis.
\\
\par
The extinction-corrected magnitude and color of the source indicate that it sits about 1.27 mag below the 
red clump centroid on the giant branch. A comparison to isochrones following \citet{bennett18,bennett_moa291} 
indicates that that source star is located on the giant branch in the Galactic bulge. Stars of similar
color and magnitude that reside in the foreground or background have a negligible probability to be lensed because of 
an extremely low number density. So, we conclude the the source star almost certainly resides in the
Galactic bulge.
\\
\par
Because the most precise determination comes from the $(V-H)$ and $H$ relation \citep{ben15}, we use the following relation to estimate $\theta_{\star}$, 
\begin{equation}
\log\theta_{\rm LD} = 0.536654 + 0.072703(V - H)_{\rm S,0} - 0.2H_{\rm S,0},
\end{equation}
where $\theta_{\rm LD}$ is the limb-darkened stellar angular diameter \citep{boy14}. This relation comes from a private communication with Boyajian by \citet{ben15}. For the best fit parameter, we get $\theta_{\star} = \theta_{\rm LD} / 2 = 3.472 \pm 0.085$ $\mu$as. 
\\
\par

\subsection{Source Angular Radius using IRSF data}
We also derive $\theta_{\star}$ using relation between $(V-K_{\rm S})_{\rm S,0}$ and $K_{\rm S,S,0}$ obtained from IRSF data. From the light curve fitting, we get $(V-K_{\rm S},K_{\rm S})_{\rm S} = (5.262,14.923) \pm (0.091,0.093)$. From the results of \citet{nis08}, we also get the extinction in $K_{\rm S}$-band $A_{\rm K_{\rm S}} = 0.222 \pm 0.005$ and color excess $E(V-K_{\rm S}) = 3.250 \pm 0.077$. Thus we find $(V-K_{\rm S}, K_{\rm S})_{\rm S,0} =  (2.011, 14.701) \pm (0.122, 0.090)$. To estimate $\theta_{\star}$, we use the following equation from \citet{ker04},
\begin{equation}
\log \theta_{\rm LD} = 0.5170 + 0.0755(V - K)_{\rm S,0} - 0.2K_{\rm S,0}.
\end{equation}
This gives $\theta_{\star} = 2.68 \pm 0.14$ $\mu$as, which is inconsistent with the one from $(V-H, H) _{S,0}$. We also get $H_{\rm S} = 15.173 \pm 0.090$ from light curve fitting of IRSF data, which is about 0.5 mag fainter than the one we get from CTIO data. This is likely because we only have three observations from IRSF for this event and our normal procedure for renormalizing error bars is not very reliable. Therefore, we adopt $\theta_{\star}$ value derived from the CTIO $V-H$ relations for the rest of the analysis. 
\\
\par

\subsection{{Color-color Relation for {\it Spitzer}}
\label{sec:colorcolor}}
\begin{figure}
\plotone{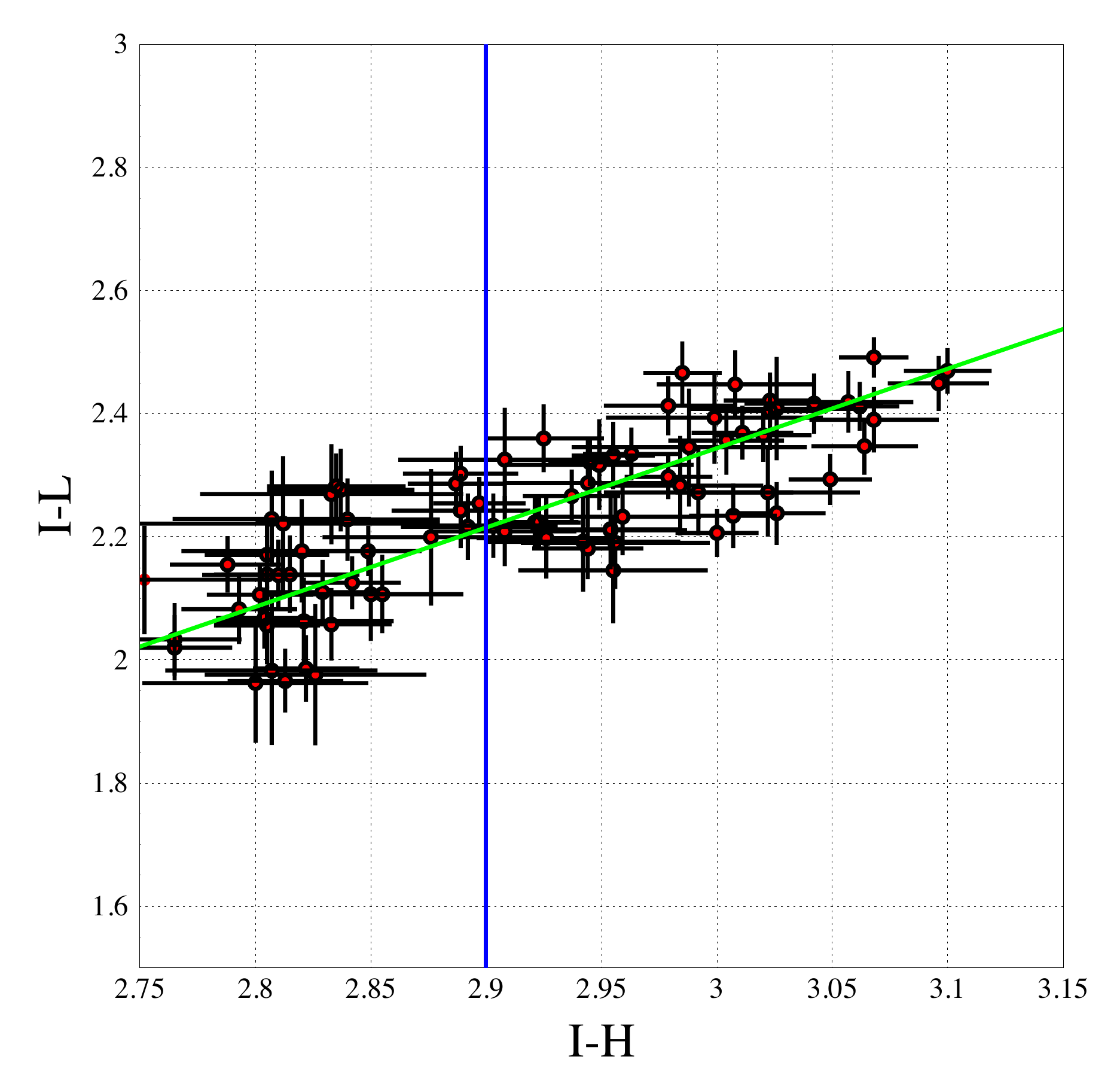}
\caption{$IHL$ color-color diagram for field stars in the neighborhood of the 
clump, $(2.75 < (I-H) < 3.10)\times (16<I<17.6)$,  The diagonal line is a fit 
to the points.  The vertical line is the observed color of the source.  The
inferred color ($y$-axis) of the source,$(I-L)_s = 2.215 \pm 0.015$ is used as
a constraint when incorporating the {\it Spitzer} data into the fit.
}
\label{fig:ihl}
\end{figure}

We construct an $IHL$ color-color diagram by matching field stars
from OGLE-IV, VVV, and our own {\it Spitzer} photometry.  We restrict
attention to stars in the neighborhood of the clump,
$(2.75 < (I-H) < 3.10)\times (16<I<17.6)$, 
and show the cross matches is Figure~\ref{fig:ihl}.
We fit these points to a straight line and find
$(I-L)= 1.289[(I-H)-2.90] + 2.215 \pm 0.008$. We find from regression 
$(I-H_{\rm CT13})_s= 1.109\pm 0.004$, and so $(I-H_{\rm VVV})_s= 2.898\pm 0.010$.
Hence this error in $(I-H)_s$ propagates to an error of 0.013 mag in $(I-L)_s$.
To this we must add in quadrature the error in the relation at the color
of the source (0.01 mag), yielding finally $(I-L) = 2.215\pm 0.015$.
\\
\par

This approach implicitly assumes that the $\sim 0.1$ mag
scatter seen in Figure~\ref{fig:ihl} is overwhelmingly due to measurement
error rather than intrinsic variation.  This is justified by 
the \citet{bb88} study of color-color relations based on bright isolated stars
with excellent photometry, which found very small scatter.
\\
\par

\section{Location and Proper Motion of the Source}
\label{sec:sourcepm}
The physical parameters that can be derived from the microlensing solution
alone appear to be quite typical of Galactic microlensing events.
That is, from $\theta_\e = 0.59\,\mas$ and $\pi_\e = 0.13$, we can
derive $M=\theta_\e/\kappa\pi_\e = 0.56\,M_\odot$ and $\pi_\rel=0.077\,\mas$,
which would be consistent with a disk lens at $D_{\rm L}\sim 5\,\kpc$
and a bulge source located at $D_{\rm S} \sim 8\,$kpc, as we have inferred 
from the source brightness and color.
This would also be consistent with the direction of lens-source
relative motion 
$\phi_\pi \equiv \tan^{-1}(\pi_{\e,E}/\pi_{\e,N}) = 32^\circ\pm 8^\circ$
(and the observed amplitude of this motion $\mu_\rel = 5.8\,\masyr$),
i.e., the direction of Galactic rotation.  This is the direction that
would be expected for a typical bulge source and a typical disk lens.
\\
\par

\subsection{{\it Gaia} Proper Motion of the Source}
\label{sec:gaiapm}
However, this seemingly clear picture appears to be contradicted by
the {\it Gaia} source proper motion 
\begin{equation}
\bmu_s(N,E) = (0.105,0.124)\pm (0.752,0.840)
\qquad (Gaia),
\label{eqn:gaia}
\end{equation}
i.e., moving synchronously with the flat disk-rotation curve, rather
than the mean motion of the bulge.  
\\
\par
The {\it Gaia} measurement is very difficult to understand within
the context of the microlensing solution.  It would imply that the
lens is moving relative to the source at 
$v_{\rm rel} \simeq \mu_\rel D_{\rm L} = 135\,\kms(D_{\rm L}/5\,\kpc)$ in the
prograde direction.  While not impossible, this would be a very
rare star.  Another alternative to consider is that $\pi_\rel$
is actually somewhat smaller (due to measurement errors of $\theta_\e$
and $\pi_\e$) so that the lens could be in the bulge.  However, the
implied motion of the lens (12 ${\rm mas\,yr^{-1}}$ relative to the mean motion of the
bulge) would be extremely rare ${\cal O}(10^{-4})$ for a bulge star.
Thus, the {\it Gaia} proper-motion measurement of the source would imply
that the otherwise quite expected microlensing parameters were either
incorrect or had extremely unusual implications.
\\
\par
This is one of two lines of argument that led us to suspect that
the {\it Gaia} measurement was actually incorrect.  The second
was that in the course of constructing a cleaned {\it Gaia} proper
motion diagram of neighboring clump stars, we noticed that stars
with parallax/error ratios $\pi/\sigma(\pi)<-2$ were preferentially
extreme proper-motion outliers.  That is, although the proper motions
of the majority of such stars were distributed similarly to those
of stars with more typical parallaxes, about 10\% had proper-motion
vectors near the edges or even outside the normal distribution and so 
were most likely to be the
result of catastrophic errors.  The {\it Gaia} parallax for the
source star is $\pi= -0.96\pm 0.42\,\mas$.  Thus, based on our
small statistical study, and even without any external reason to suspect
the measurement, the strong negative parallax implied a $\sim 10\%$ 
probability of a catastrophic error.  We also note that this star has
an ``astrometric excess noise sig'' of 3.19.  However, we show below
that this is actually substantially below the median of a well-behaved
``clean clump and near-clump sample''.  So this value is not, in itself,
a reason to be suspicious of this star.
\\
\par

\subsection{OGLE Proper Motion of the Source}
\label{sec:oglepm}
\begin{figure}
\plotone{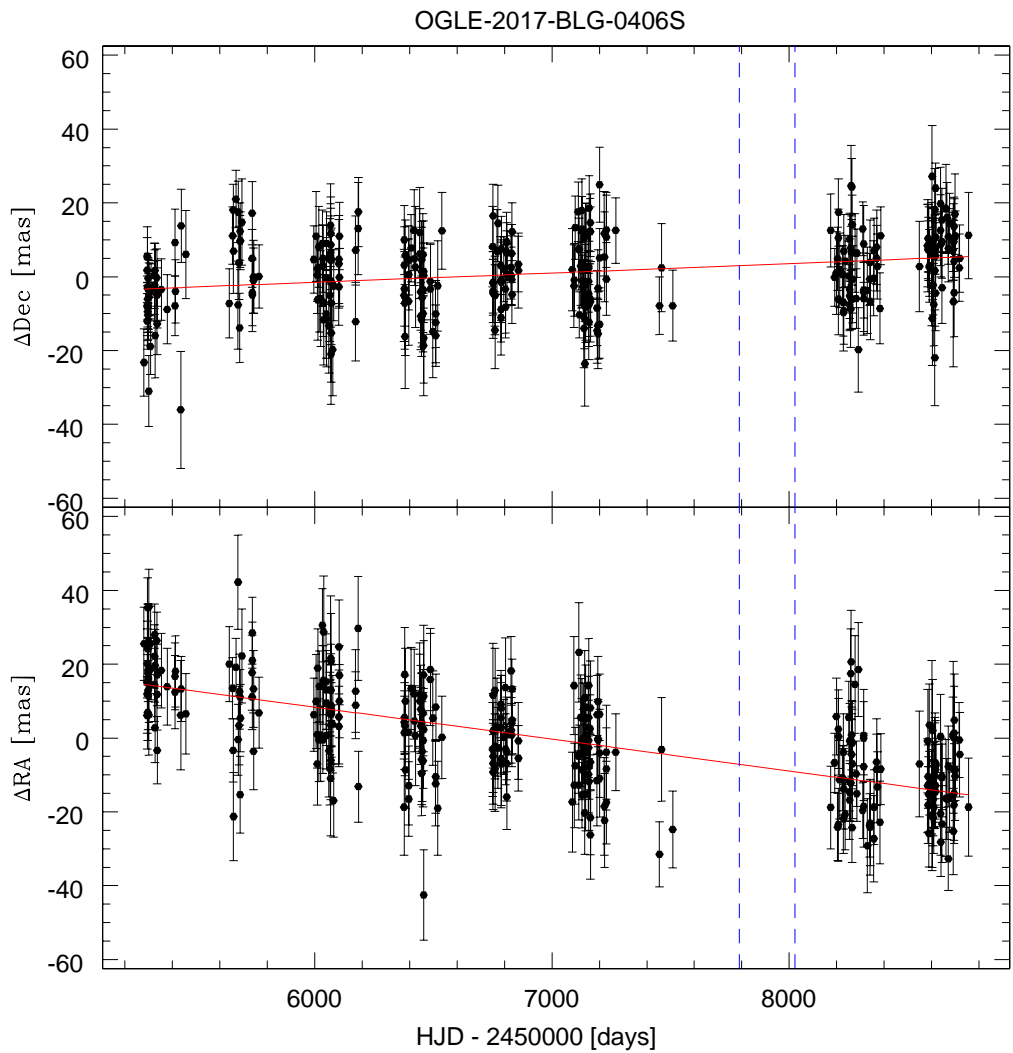}
\caption{Individual position measurements converted to mas from  OGLE-IV pixels
($0.26^{\prime\prime}$) on the $y$ axis (north, upper) and negative-$x$ axis 
(east, lower) of the detector.
The observed slopes are $0.92\pm 0.16\,\masyr$ (north) and 
$-3.15\pm 0.18\,\masyr$ (east).
By contrast, the measurements reported by {\it Gaia} would yield
corresponding slopes of
$5.57\pm 0.79\,\masyr$ and $3.52\pm 0.84\,\masyr$
, respectively.  The {\it Gaia}
measurement is therefore directly contradicted by the OGLE data.
}
\label{fig:oglepm}
\end{figure}

There is a long history of OGLE proper measurements of bulge sources
dating back to the \citet{sumiogle} catalog based on OGLE-II data.  While
there are no published catalogs based on the subsequent OGLE
surveys, individual proper-motion measurements based on OGLE-IV are
potentially more precise by a factor of several tens due to a five-times
longer baseline and equal or higher cadence 
\citep{mb11028,ob161540,mb16231,ob170896}.  
We apply this same technique to the OGLE-2017-BLG-0406
source and find, in the OGLE-IV reference frame tied to 1050 red clump stars
within a $(6.5^\prime \times 6.5^\prime)$ square,
$\bmu_{s,{\rm OGLE-IV}}(N,E) = (0.923,-3.147)\pm (0.163,0.182)\,\masyr$,
where the errors are derived by assuming that errors of the individual
position measurements are equal to the rms scatter about the best-fit
straight line, i.e., $\sigma(N,E) = (10,11)\,\mas$, which corresponds to
about 0.04 OGLE pixels.
Figure~\ref{fig:oglepm} shows the 324 data points
during the period 2010-2019 that went into this measurement.
\\
\par

We then align the local OGLE-IV proper motion frame to the {\it Gaia}
frame by cross matching common stars.  For this purpose, we consider
{\it Gaia} stars within $\Delta\theta<3^\prime$ and with
``astrometric excess noise sig'' $<10$ and then further restrict
to our ``clean clump
and near-clump sample'', which is defined by
$16<G<18$, $2<(B_P-R_P)<3$, 
$\sigma(\mu_{\rm RA})<0.6\,\masyr$, $\sigma(\mu_{\rm Dec})<0.6\,\masyr$,
$\pi/\sigma(\pi)>-2$, and $\pi<1\,\mas$.  For purposes of finding the
offset between two proper motion frames, there is no reason to restrict
to clump stars.  However, the clump (and near-clump) sample allows us
to identify and reject several data classes that are prone to catastrophic
errors.  We find
$\Delta\bmu(N,E)=\bmu_{Gaia}-\bmu_{\rm OGLE-IV}
= (-5.552,-3.391)\pm(0.042,0.052)\,\masyr$,
based on an initial set of 394 stars from which we eliminate
nine and six three-sigma outliers, respectively.
Hence, we obtain 
$\bmu_{s,Gaia}=\bmu_{s,{\rm OGLE-IV}}+\Delta\bmu = (-4.63,-6.54)\pm (0.17,0.19)$.
\\
\par
While these small formal errors accurately reflect the OGLE-IV measurement
of the ``catalog star'' associated with the microlensed source,
this catalog star is composed of both the source and a very small amount
of blended light.  Subtracting the precisely determined source flux
from the somewhat more uncertain baseline flux of the catalog star,
this blended flux is about 7\% of the total.  The true number could
be slightly more or less.  To take account of the possibly different
proper motion of the blend, we augment the error in the source proper
motion to a somewhat conservative $0.4\,\masyr$,
\begin{equation}
\bmu_s(N,E) = (-4.63,-6.54)\pm (0.40,0.40).
\qquad ({\rm OGLE-IV}),
\label{eqn:ogleiv}
\end{equation}
Note that Equations~(\ref{eqn:gaia}) and (\ref{eqn:ogleiv}) differ by
about $8.1\,\masyr$ or about $10\,\,\sigma$ using the reported {\it Gaia}
uncertainties.
\\
\par
Our threshold of ``astrometric excess noise sig'' may appear at first
site to be too generous.  However, we find that in our final sample of
394 OGLE-{\it Gaia} matches, a fraction $(6,15,30,48)\%$ lie below
$(1,2,3,4)$ respectively, with a median of 4.1. Yet, the sample as whole
has well behaved proper motions, with only 1--2\% three-sigma outliers
relative to OGLE.
\\
\par

\begin{figure}
\plotone{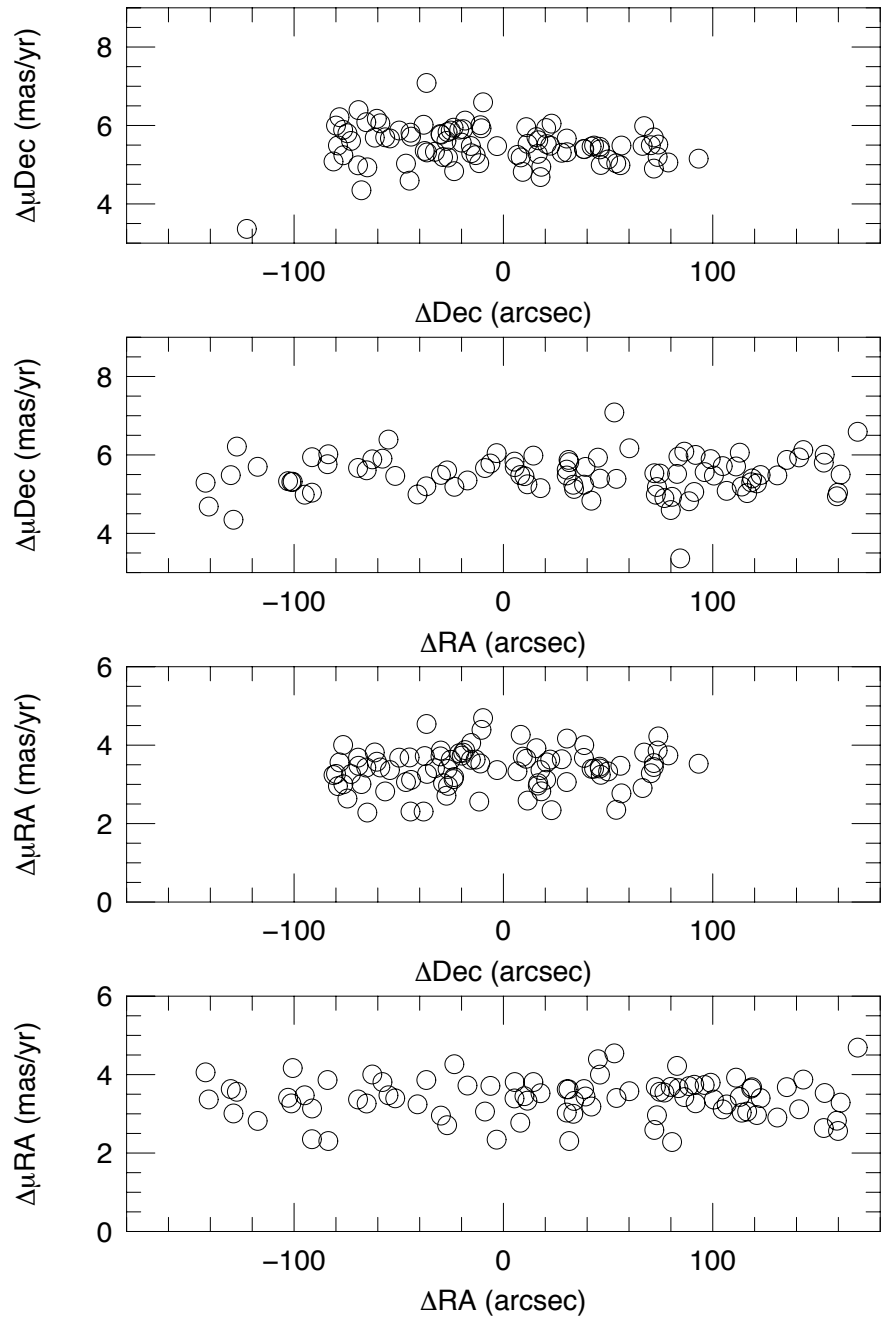}
\caption{Offset of OGLE-IV proper motions relative to {\it Gaia}
as a function of position in Equatorial coordinates (as indicated).
}
\label{fig:offsets}
\end{figure}

To find the offset $\Delta\bmu$ between the {\it Gaia} and OGLE-IV
systems, we fit the proper-motion differences to a quadratic function
of position centered at the lensed-source position.  Because the
formal {\it Gaia} errors were several times larger than the formal
OGLE-IV errors, we considered only the former, and we rescaled these
errors to enforce $\chi^2/{\rm dof} =1$.  This yielded rescaling
factors of 2.22 and 2.14 in the north and east directions, respectively.
Figure~\ref{fig:offsets} shows the proper-motion offsets as a function of 
each Equatorial coordinate.  This figure shows some large-scale structure,
which is removed by the quadratic fits, as well as some small-scale
structure, which is not.  However, this small-scale structure is
relatively isolated and has a amplitude of a few tenths $\masyr$, so
it is unlikely to account for the increased scatter, which is of order
$1\,\masyr$.  Rather, the most likely source of the majority of this additional
scatter is underestimation of {\it Gaia} errors, which is exactly
what is corrected by our error-renormalization procedure.  To further explore
this idea, for each star in our ``clean clump and near-clump sample'' (but
now re-including the stars with $\pi/\sigma(\pi)<-2$), we calculate
a parallax offset parameter $\eta=(\pi-\pi_0)/\sigma(\pi)$, where
$\pi_0 =\pi_{\rm bulge} - \pi_{\rm zpt}= 70\,\muas$, and
we adopt $\pi_{\rm bulge} = 120\,\muas$ for the mean parallax of the clump and
$\pi_{\rm zpt} = 50\,\muas$ for zero-point offset of the Gaia parallax system.
After restricting attention to $|\eta|<4$ we find
$\langle\eta\rangle = -0.43$ (compared to zero expected),
$\sigma(\eta) = 1.31$ (compared to unity expected) and 
$\sqrt{\langle \eta^2\rangle} = 1.38$ (compared to unity expected).
If the error properties of the proper motions were similar to those of
the parallaxes (as would be expected) then these numbers would partially
explain the higher-than-expected scatter in the {\it Gaia}-OGLE-IV 
comparison.
\\
\par
One possible source of this {\it Gaia} astrometry error is blending in the 
crowded Galactic bulge field. {\it Gaia} has an asymmetric PSF that can 
lead to blending with another star at a separation of $\sim 0.15^{\prime\prime}$ in some 
passes and not others. Such a circumstance would likely invalidate the {\it Gaia} 
astrometry, which could lead to negative parallaxes and proper motion errors.
\\
\par

\section{{Physical Parameters}
\label{sec:physical}}
\begin{figure}
\begin{center}
\includegraphics[width=.4\linewidth]{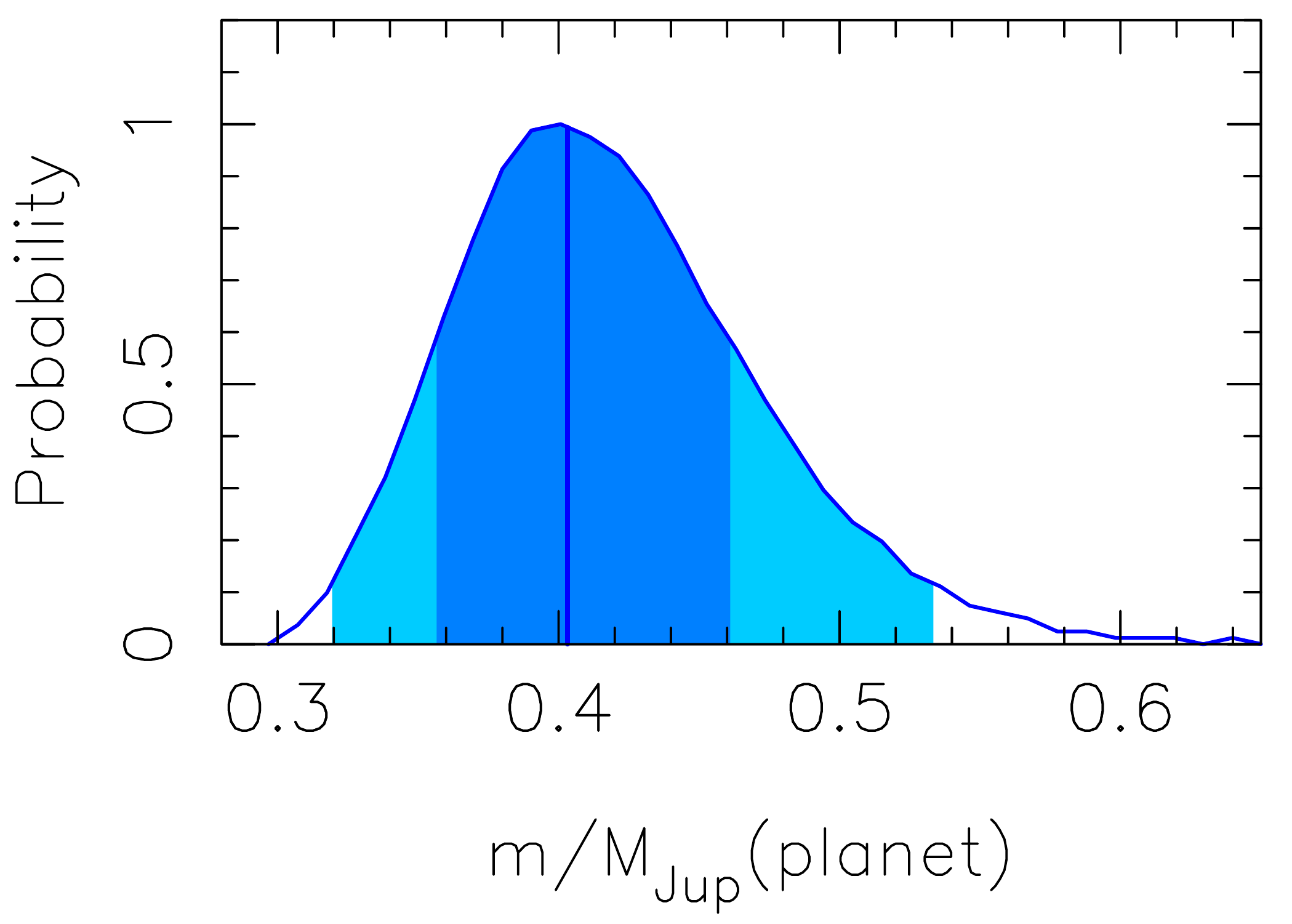}\quad\includegraphics[width=.4\linewidth]{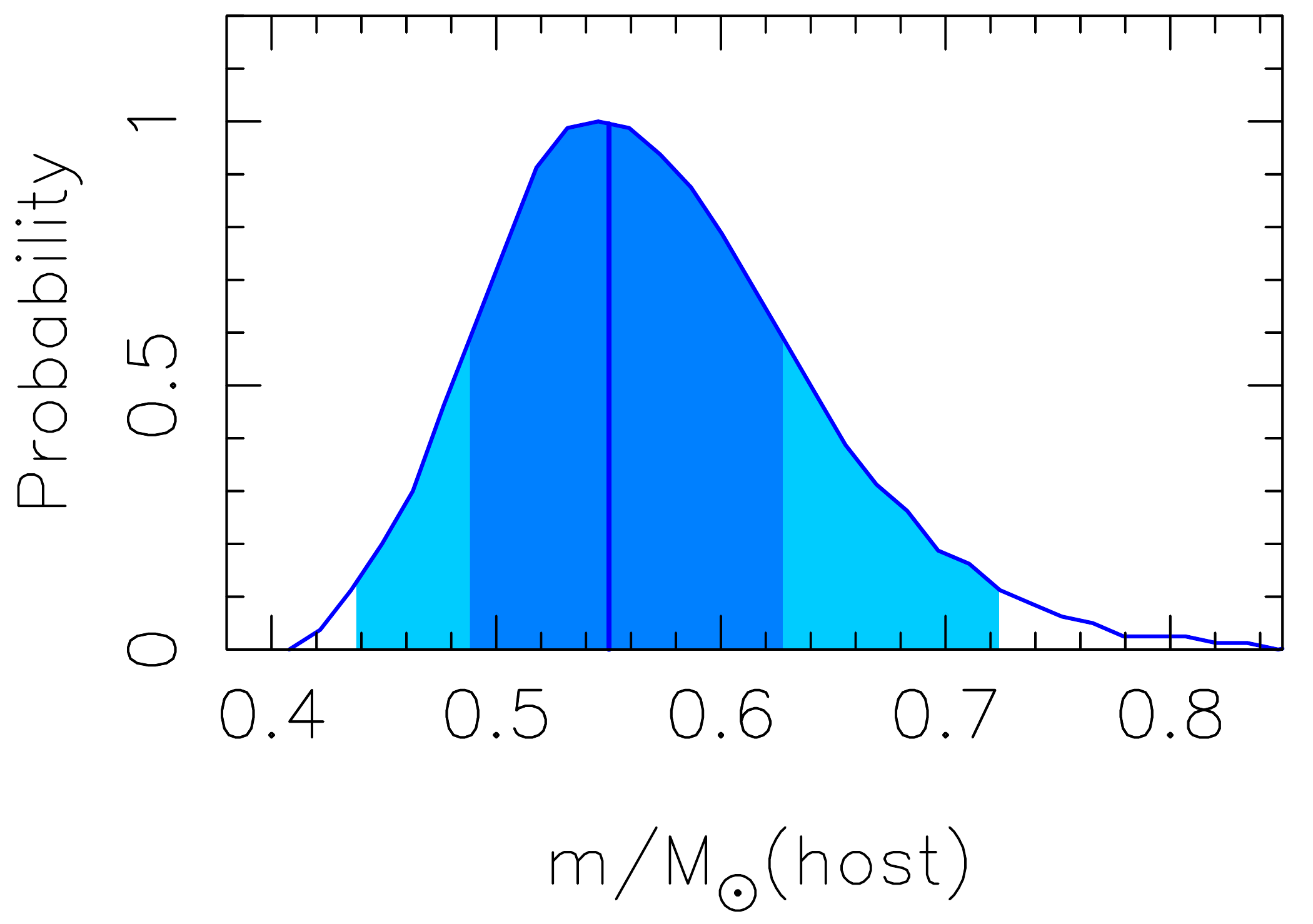}
\\[\baselineskip]
\includegraphics[width=.4\linewidth]{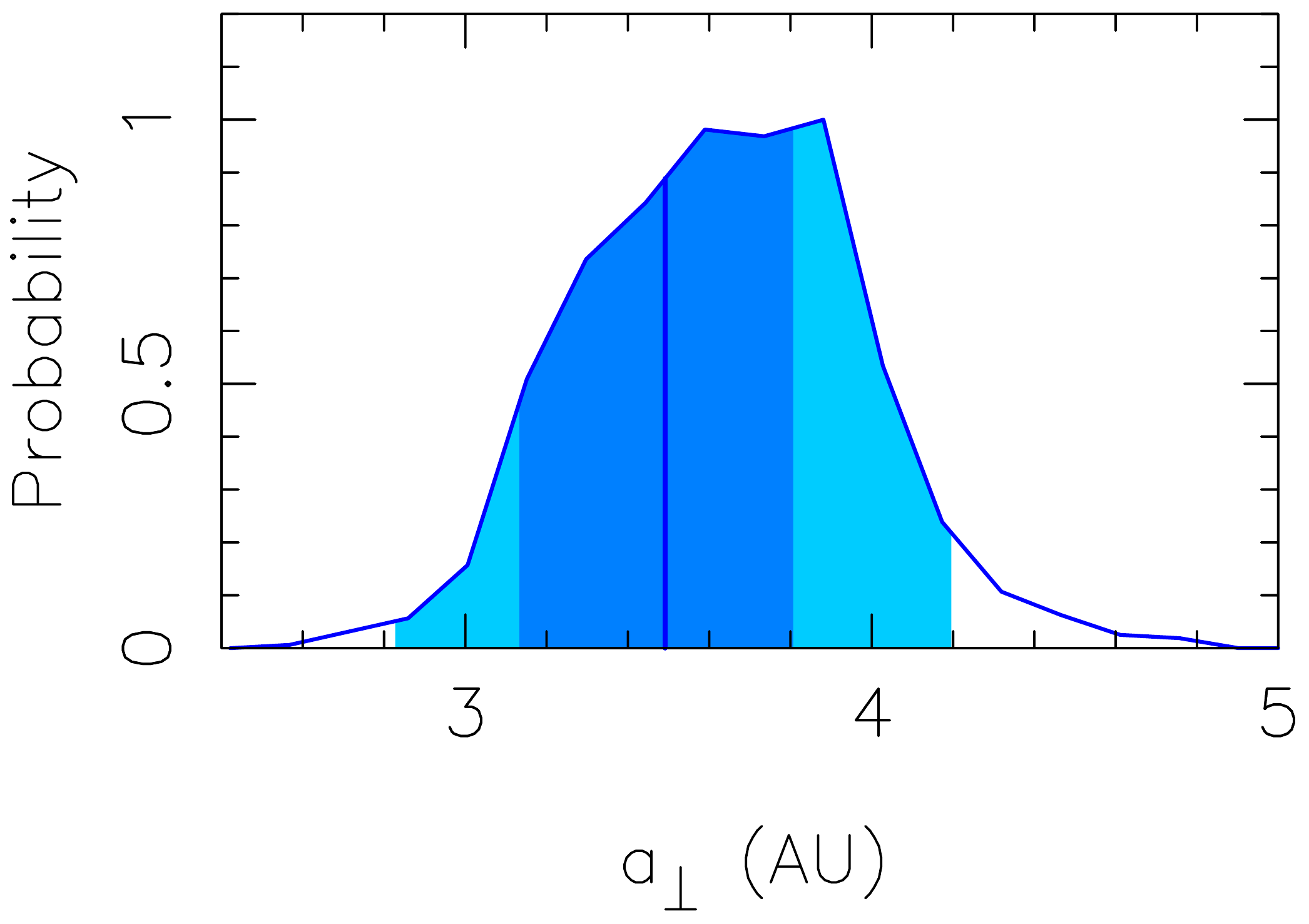}\quad\includegraphics[width=.4\linewidth]{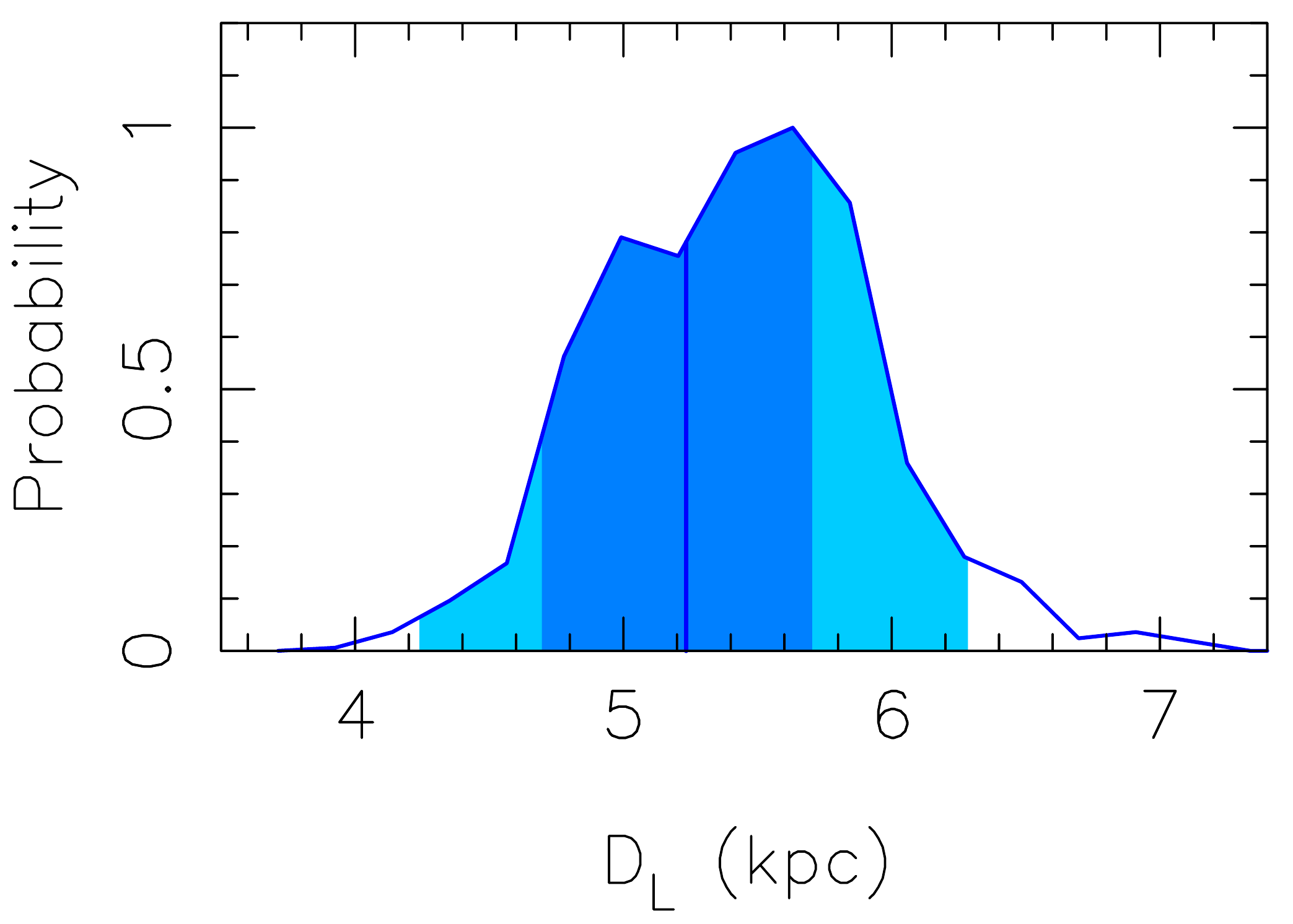}
\caption{Probability distributions of lens properties of planetary mass, $M_{\rm planet}$, host star mass, $M_{\rm host}$, projected separation, $a_{\perp}$, and distance, $D_{\rm L}$ from our Bayesian analtsis. The dark and light blue regions indicate the 68.3\% and 95.4\% confidence intervals, and the vertical blue lines indicate the median value.\label{fig:lens-prop}}
\end{center}
\end{figure}

\begin{figure}
\begin{center}
\includegraphics[width=.4\linewidth]{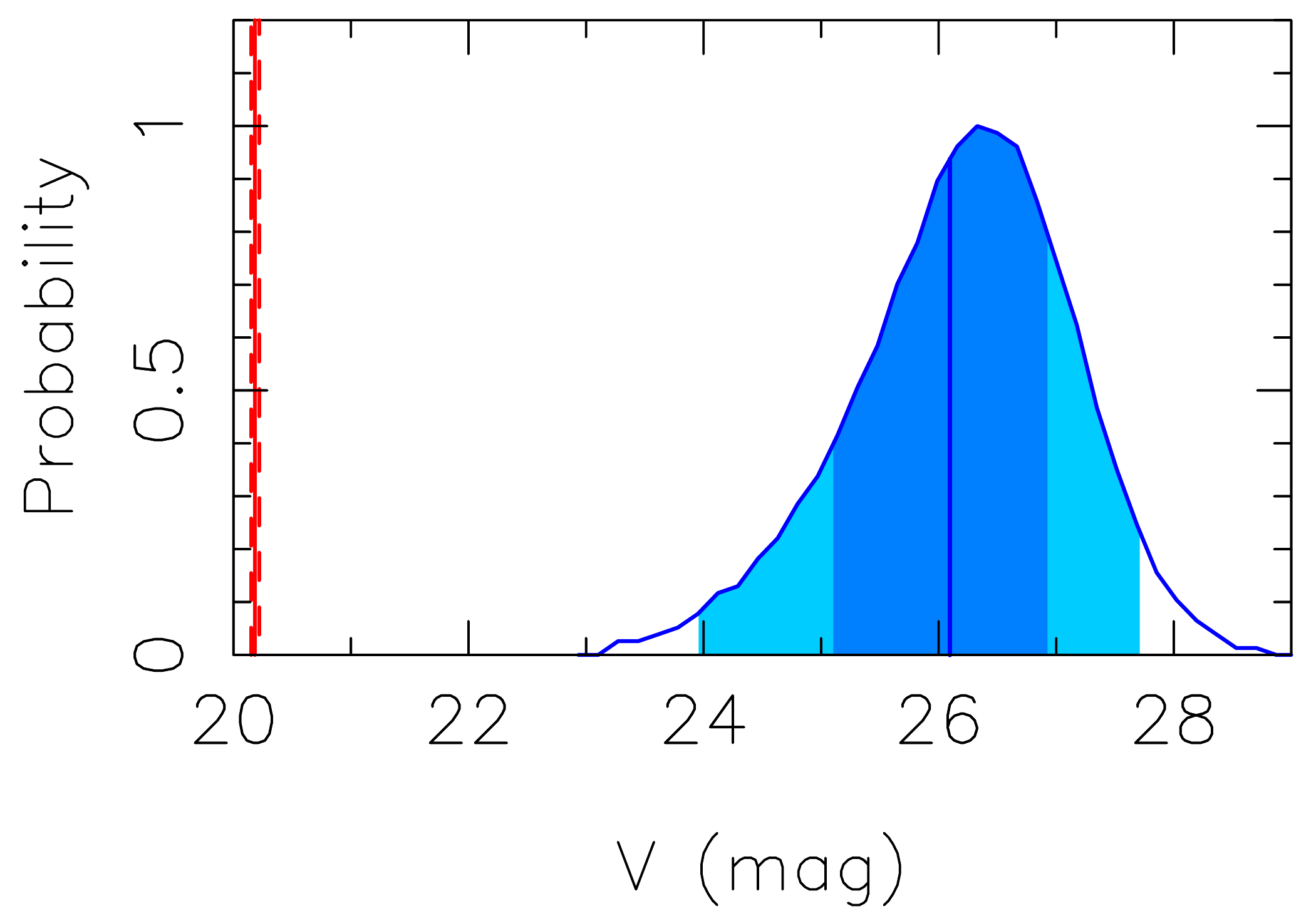}\quad\includegraphics[width=.4\linewidth]{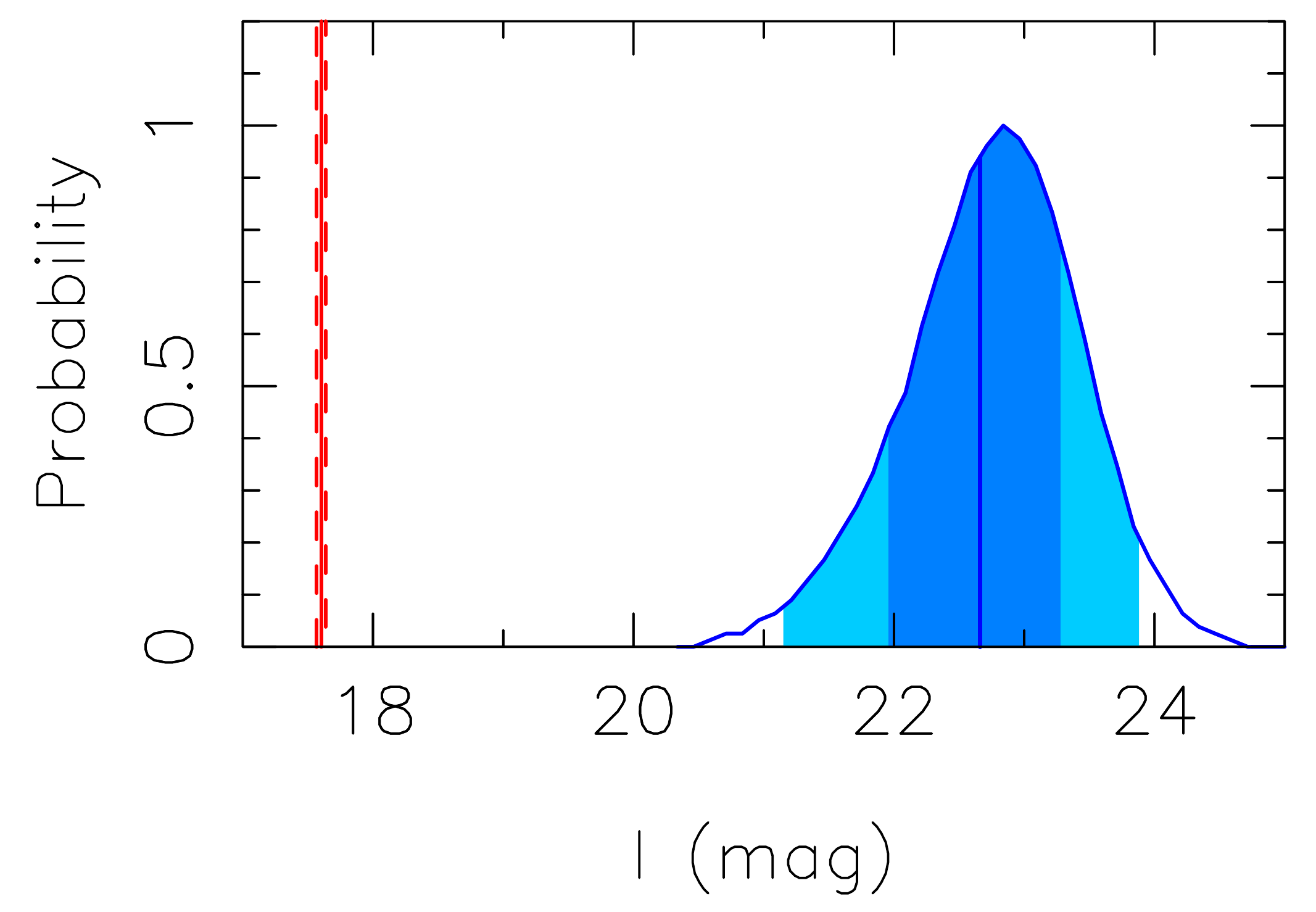}
\\[\baselineskip]
\includegraphics[width=.4\linewidth]{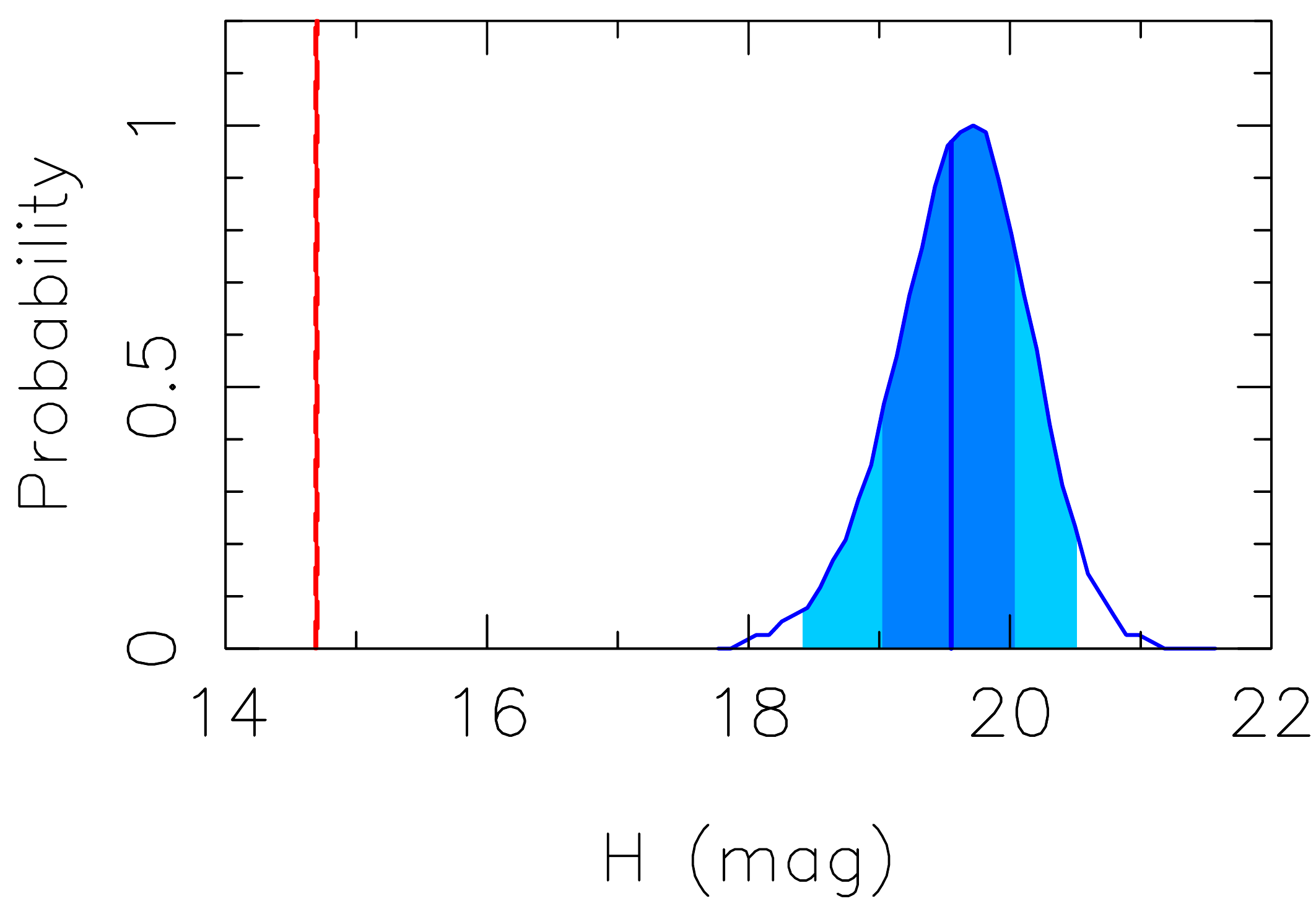}
\caption{Probability distributions of lens brightness with extinction. The dark and light blue regions indicate the 68.3\% and 95.4\% confidence intervals, and the vertical blue lines indicate the median value. The red solid and dashed lines indicate the source brightness and its 1 $\sigma$ errors from the light curve fitting \label{fig:lens-mag}}
\end{center}
\end{figure}

Because the microlens parallax vector, $\bpi_\e$, the amplitude of
the lens-source relative proper motion, $\mu_\rel$, and the source
proper motion, $\bmu_s$, are all well measured, and
the source distance, $D_{\rm S}$, is constrained to reside in the Galactic bulge, 
we can directly calculate the lens physical parameters, namely
\begin{equation}
M_{\rm host} ={\theta_\e\over(1+q)\kappa\pi_\e},
\qquad
\pi_\rel =\theta_\e\pi_\e,
\qquad
\bmu_\rel = {\bpi_\e\over\pi_\e}\mu_\rel,
\qquad
(\pi,\bmu)_{\rm L} = (\pi,\bmu)_{\rm S} + (\pi,\bmu)_\rel.
\footnote{$\bmu_{\rm rel}$, $\bmu_{\rm L}$ and $\bmu_{\rm S}$ are in geocentric coordinate.} 
\label{eqn:phys}
\end{equation}
We compute these quantities, as well as $M_{\rm planet}$,
$\bmu_{\rm rel,H}$, $a_\perp$, and ${\bf v}_{\rm L}$
from the MCMC, using a Galactic prior, 
and we report
the results in Table~\ref{tab:phys} and Figures~\ref{fig:lens-prop} and \ref{fig:lens-mag}.  
The host star mass is denoted by $M_{\rm host}$, and the 
planet mass is given by $M_{\rm planet} = qM_{\rm host}$.
\\
\par

\begin{deluxetable}{lcl}
\tablecaption{Lens Physical Parameters \label{tab:phys}}
\tablehead{
\colhead{Parameter} & \colhead{units} & \colhead{Values} 
}
\startdata
$M_{\rm host}$ & $M_{\odot}$ & $0.56\pm  0.07$  \\
$M_{\rm planet}$ & $M_{\rm Jup}$ & $0.41\pm  0.05$ \\
$D_{\rm S}$  & kpc & $8.8\pm 1.2$ \\
$D_{\rm L}$  & kpc &  $5.2\pm 0.5$ \\
$a_{\perp}$ & au & $3.5\pm 0.3$ \\
$a_{\rm 3d}$ & au & $4.1^{+2.1}_{-0.7}$ \\
$\theta_{\rm E}$ & mas & $0.593 \pm 0.012$ \\
$\mu_{\rm rel}$ & mas/yr & $ 5.84 \pm 0.12$ \\
$\mu_{\rm rel,H,N}$  & mas/yr     & $  5.1\pm  1.0$  \\
$\mu_{\rm rel,H,E}$  & mas/yr      & $  3.39\pm  0.37$ \\
$v_{{\rm L},l}$  &  km/sec   & $   230\pm    33$ \\
$v_{{\rm L},b}$ &  km/sec  & $    64\pm    8$ \\
$V_{\rm S}$ & mag    &    $   20.187 \pm 0.020  $  \\  
$I_{\rm S}$ & mag     &    $ 17.606 \pm 0.020   $ \\
$H_{\rm S}$ & mag    &    $  14.697 \pm 0.020  $ \\
$V_{\rm L}$ & mag    &    $   26.1  \pm 0.9  $  \\
$I_{\rm L}$ & mag   &    $   22.7  \pm  0.7 $ \\
$H_{\rm L}$ & mag   &    $   19.6 \pm 0.5   $ \\
\enddata
\end{deluxetable}

We use a different Galactic prior than previous microlensing analyses \citep{sum11,ben14,zhu17} because
it is now clear that these older model have several incorrect features. 
One such feature is the varying Galactic disk velocity dispersion as a function of the Galactocentric distance, $R$, which
increases when $R$ gets smaller, while the disk scale height decreases with decreasing $R$.
Another important feature is the changing distribution of the azimuthal velocity $V_{\phi}$
as a function of Galactocentric distance, $R$. If we define the circular velocity $V_c(R)$ as the
velocity of a circular orbit at a distance $R$, then we find that 
there are more stars
with $V_{\phi} < V_c$ than stars with $V_{\phi} > V_c$ at a given Galactocentric distance $R$,
where $V_c$ is the circular velocity at $R$.
Both of these features are observed in the {\it Gaia} DR2 data \citep{kat18}.
Koshimoto et al. (in preparation) developed a Galactic model that is based on
the Shu distribution function model by \citet{sha14}, but modified so that
the mean velocity and velocity dispersion as a function of the Galactocentric distance $R$ and
the height from the Galactic plane $z$ match the {\it Gaia} DR2 data \citep{kat18}. Table~\ref{tab:V5kpc}
summarizes the distribution of Galactic transverse velocities in this Galactic model. It gives the
median and 1 and $2\sigma$ values for the stellar velocities for thin disk, thick disk, bulge and all stars (i.e., the median and 15.85, 84.15, 2.28, and 97.72
percentiles of the transverse velocity distribution). The Galactic circular velocity in the Solar neighborhood is
$V_c = 238.8\,$km/sec in our model.
We used this model as the Galactic prior to calculate the lens properties. 
\\
\par

\begin{deluxetable}{ccccccc}
\tablecaption{Model Stellar Velocities at $D = 5.2\pm 0.5$ \label{tab:V5kpc}}
\tablehead{
\colhead{Star Component} & \colhead{Velocity Component}  & & & & &  \\
\colhead{} & \colhead{km/sec} & \colhead{$-2\sigma$} & \colhead{$-1\sigma$}  & \colhead{median ($\widetilde{v}$)} &  \colhead{$+1\sigma$} & \colhead{$+2\sigma$}
}
\startdata
Thin Disk stars & $v_l$ & 92.4 & 148.3 & 195.1 & 236.2 & 278.4 \\
& $v_b$ & -70.4 & -30.9 & 0.8 & 30.9 & 70.2 \\
Thick Disk stars & $v_l$ & 54.7 & 118.0 & 182.0 & 245.2 & 312.7 \\
& $v_b$ & -127.3 & -63.9 & 1.1 & 67.5 & 127.8 \\
Bulge stars & $v_l$ & -12.6 & 50.8 & 112.3 & 172.7 & 231.2 \\
& $v_b$ & -109.6 & -53.8 & -0.4 & 54.6 & 112.5 \\
All Stars & $v_l$ & 37.6 & 120.6 & 185.0 & 231.7 & 277.9 \\
& $v_b$ & -86.3& -36.1 & 0.6 & 36.3 & 86.8 \\
\enddata
\end{deluxetable}

We can determine the heliocentric proper motion of the lens, $\bmu_{\rm L,H}$,
because we have measured the microlensing parallax vector, $\bpi_{\rm E}$, using
\begin{equation}
\bmu_{\rm L,H} \equiv \bmu_{\rm L} + {\pi_{L}\over au}{\bv_{\oplus,\perp}};
\qquad
\bv_{\oplus,\perp}(N,E) = (+0.69,+28.27)\,\kms\ ,
\label{eqn:bmuhel}
\end{equation}
where $\bv_{\oplus,\perp}$ is the projected velocity of Earth at $t_{o,\oplus}$.
We can determine the velocity of the lens system with
\begin{equation}
\bv_{\rm L} = D_{\rm L} \bmu_{\rm L,H} + \bv_{\odot,\perp}
\qquad
\bv_{\odot,\perp}(l,b) = (+250.8,+7)\,\kms,
\label{eqn:vlsr}
\end{equation}
and $\bv_{\odot}$ is 
the motion of the Sun in the Galactic frame, which includes
its peculiar motion relative to the circular velocity, 
$V_c(R_0)$, at the solar circle, $R_0$, in a Galaxy-centered coordinate system. 
The lens velocity, $\bv_{\rm L}  = (v_{{\rm L},l}, v_{{\rm L},b})$,
can then be compared with the median velocity, ($\widetilde{v_l}$, $\widetilde{v_b}$), of the
Galactic stars at the distance of the lens system, $D_{\rm L} = 5.2\pm 0.5$ kpc, as indicated
in Table~\ref{tab:V5kpc}.
\\
\par
Table~\ref{tab:phys} indicates that the system is composed of an early 
M-dwarf host
($M_{\rm host} = 0.56\pm 0.07\, M_\odot$) orbited by a Saturn-mass planet
($M_{\rm planet} = 0.41\pm 0.05\, M_{\rm Jup}$) at projected separation,
$a_\perp = 3.5\pm 0.3\,\au$, i.e., just over twice the snow line
(assuming that this scales as $r_{\rm snow} = 2.7\,\au(M/M_\odot$).
The lens system lies at $D_{\rm L} = 5.2\pm 0.5\,\kpc$, i.e., somewhat more
than halfway toward the bulge.  It is moving in the azimuthal direction at a speed, $v_{{\rm L},l} = 230\pm 33\,$km/sec,
that is just $1\sigma$ above the median ($\widetilde{v_l} = 195\,$km/sec or 182\,km/sec) for thin and thick disk stars, 
and $2\sigma$ above the median ($\widetilde{v_l} = 112.3\,$km/sec) for bulge stars, given
in Table~\ref{tab:V5kpc}. The lens
vertical velocity $v_{{\rm L},b} =64\pm 8\,$km/sec is within $1\sigma$ above the median for thick disk stars and 
between $1\sigma$ and $2\sigma$ above the
median for the thin disk and bulge stars. The thin disk, thick disk and bulge stars comprise 80\%, 11\%, and
9\%, respectively, of the stars that provide $t_{\rm E}\approx 37\,$days events at $D_{\rm L} = 5.2\pm 0.5\,$kpc.
Therefore, the lens system is most likely to be part of the thin or thick disk population, but a bulge lens
system cannot be ruled out.
\\
\par
The source and lens magnitudes are also given in Table~\ref{tab:phys} and Figure~\ref{fig:lens-mag}. The lens
magnitudes were calculated from the host star masses from our MCMC over light curve models using the
empirical mass-luminosity relation described by \citet{bennett_moa291}, which is a combination of
several different mass-luminosity relations for different mass ranges. 
For $M_L \geq 0.66\,\msun$, $0.54\,\msun\geq M_L \geq 0.12\,\msun $, and 
$0.10 \,\msun \geq M_L \geq 0.07\,\msun$, we use the relations of \citet{henry93}, \citet{delfosse00},
and \citet{henry99}, respectively. In between these
mass ranges, we linearly interpolate between the two relations used on the
boundaries. That is, we interpolate between the \citet{henry93} and the \citet{delfosse00}
relations for $0.66\,\msun > M_L > 0.54\,\msun$, and we interpolate between the
\citet{delfosse00} and \citet{henry99} relations for $0.12\,\msun > M_L > 0.10\,\msun$.
\\
\par
The detection of the lens star in follow-up observations will be somewhat challenging 
because the source is a first-ascent giant only 1.3 magnitudes fainter than the red clump.
The median predicted lens magnitude, $H_{\rm L}$, is 4.9 magnitudes fainter than the source,
which means that it is fainter than the calibration uncertainty in $H_{\rm L}$, so we cannot
expect to detect any significant excess flux at the position of the source, unless the lens
is near the $2\sigma$ upper limit on its brightness. Also, the relatively red source implies that
it will be difficult to detect the lens star using the color-dependent centroid shift \citep{ben06} 
because the lens is likely to have a similar color to the source. The image elongation \citep{ben07} is 
also difficult to measure with such a high ratio between the source and lens brightnesses.
Thus, the detection of the lens star will be significantly more
challenging than previous cases with Keck adaptive optics (AO) imaging
\citep{bat15,bha18,van19,ben20}, in which image separations of 0.62 to 1.47 FWHM
were measured at flux ratios of 1.46 to 3.15.  Extrapolating from
Figure~1 of \citet{ben20}, we estimate that the lens and source
can be confidently resolved at 1.3 FWHM, i.e., 72 mas with Keck
$K$-band AO.  This requires waiting until 2029.  In any case, they
would be resolved at first AO-light on the next generation of
extremely large telescopes (ELTs).
\\
\par

\section{{Discussion}
\label{sec:discuss}}

\subsection{{Eighth {\it Spitzer}-sample Planet}
\label{sec:spitzsamp}}
In order to derive statistically robust conclusions about planets 
from a given microlensing sample, the events must enter the sample
without regard to whether they have planets or 
not\footnote{But see \citet{ob171434} for an alternative planet-only
based approach.}.  For the {\it Spitzer} sample, this criterion
gives rise to two distinct issues: (1) the events should be chosen
for observations and assigned an observational cadence 
without regard to the presence of planets, and (2)
the data-quality threshold for entering the sample should be the
same for events with and without planets.
\\
\par
Regarding the first, \citet{yee15} gave detailed prescriptions for
including events under various modes of selection.  However, from
the present perspective the situation is relatively simple: 
OGLE-2017-BLG-0406 met so-called ``objective'' criteria and thus
had to be observed regardless of whether it had a planet or not.
Moreover, it was observed at an ``objectively determined'' cadence.
The only exception to this was that, as with essentially all known
planetary events, OGLE-2017-BLG-0406 was observed at baseline during
the 2019 season.  The main motivation for this was to test for
systematics, in part due to concerns raised by \citet{kb19}.  
See, for example, \citet{kb180029}.
The addition of baseline data leads to a more precise parallax measurement.
Thus, while these additional data are not in themselves relevant to the 
present point, they are to the next one.
\\
\par
Second, \citet{zhu17} established the following data-quality condition
for the statistical study of the Galactic distribution of planets:
the parallax should be adequately measured, meaning that the event should 
only be
included in the sample provided that the error in ``$D_{8.3}$'' satisfies
\begin{equation}
\sigma(D_{8.3})<1.4\,\kpc;
\qquad
D_{8.3} \equiv {\kpc\over \pi_\rel/\mas + 1/8.3},
\label{eqn:d8point3}
\end{equation}
where $\pi_\rel = \theta_\e\pi_\e$.
However, in order that this criterion be independent of the presence of the
planet,  they require that the estimate of $\sigma(D_{8.3})$
be derived from the corresponding single-lens event (with planet removed), 
rather than the actual event (which has added information from the
planetary anomaly and (possibly) finite-source effects. 
\\
\par
Therefore, we tested whether this event
meets this criterion as follows. First, we make the analogous
data set as described in \citet{ob161190}. Next, we fit the data set
with a single lens model with parallax and finite-source effects, i.e.,
six microlensing parameters $(t_o,u_0,t_\e,\rho,\pi_{\e,N},\pi_{\e,E})$.
Note that for this purpose we remove the 2019 {\it Spitzer} data, 
because these would not have been obtained if there had been no planet.
\\
\par
In typical cases of single-lens events (whether real or, as in this
case, simulated), one does not measure $\rho$, but only obtains some
(usually weak) limit $z_0\gg 1$ where $z_0\equiv u_0/\rho$.  Thus, one
must estimate $D_{8.3}$ using a Bayesian analysis \citep{zhu17}.  
In the present case,
$z_0 = 1.59$.  Because $z_0>1$, there is no caustic crossing in the
single-lens event, but there is significant excess magnification at
peak relative to the point-source case, which can be evaluated in the
quadrupole approximation \citep{gould08} as
\begin{equation}
{\delta A\over A} = {1 - \Gamma/5\over 8 z_0^2} \rightarrow 4.5\% .
\label{eqn:quadrupole}
\end{equation}
Here we have adopted $\Gamma = 0.5$ as an illustrative value of
the limb-darkening parameter.  Therefore given the high density and
precision of the data over the peak, as well as over most of the event,
we expect a very good measurement of $\rho$.  In fact we find that
the single-lens light curve yields excellent constraints on both
$\bpi_\e$ and $\rho$, with $\sigma(D_{8.3})= 0.19\,\kpc$
(compared to $0.26\,\kpc$ for the actual event).  Therefore,
in this case, it is not necessary to conduct additional Bayesian analysis
because the single-lens events satisfies the \citet{zhu17} criterion
without it.
\\
\par
Hence, OGLE-2017-BLG-0406Lb becomes the eighth planet in the {\it Spitzer}
statistical sample, the others being
OGLE-2014-BLG-0124Lb \citep{ob140124}
OGLE-2015-BLG-0966Lb \citep{ob150966},
OGLE-2016-BLG-1190Lb \citep{ob161190}, 
OGLE-2016-BLG-1195Lb \citep{ob161195a,ob161195b},
OGLE-2017-BLG-1140Lb \citep{ob171140}, 
OGLE-2018-BLG-0799Lb \citep{ob180799},   and
KMT-2018-BLG-0029Lb \citep{kb180029}.
In addition, there are two microlensing planets from the {\it Spitzer}
bulge survey that do not enter the statistical sample,
OGLE-2016-BLG-1067Lb  \citep{ob161067} and
OGLE-2018-BLG-0596Lb  \citep{ob180596},
as well as one other {\it Spitzer} microlensing planet in the Galactic disk,
Kojima-1b \citep{nucita18,fukui19,kojima2}. 
To our knowledge, there are two other potential 
{\it Spitzer}-statistical-sample planets under active investigation.
The {\it Spitzer} microlensing program ended in 2019.
\\
\par

\subsection{{1-D {\it Spitzer}-``only'' Parallax $\cap$ 1-D Ground-only Parallax}
\label{sec:cap}}
Figures~\ref{fig:arc+} and \ref{fig:arc-} show that the OGLE-2017-BLG-0406
parallax measurement derives from the intersection of two sets of
1-D parallax contours, one from the ground-only measurement and the
other from the {\it Spitzer}-``only'' measurement.  \citet{gould99}
first suggested the idea of combining 1-D parallax information from
{\it Spitzer} with 1-D information from the ground, but after {\it Spitzer}
observations of $\sim 1000$ microlensing events, this is only the
second case for which the intersection of such 1-D information has been
demonstrated.
\\
\par
However, this lack of identified cases may simply reflect the fact
that OGLE-2017-BLG-0406 is only the fourth microlensing event for which
{\it Spitzer}-``only'' and ground-only parallax contours have been
shown separately\footnote{\citet{ob180596} separately analyzed the
{\it Spitzer}-``only'' and ground-only $\bpi_\e$ measurements for
OGLE-2018-BLG0596, but they only showed contours for the former.  We
have checked, using an analog of Figures~\ref{fig:arc+} and \ref{fig:arc-}, that
the $1\,\sigma$ parallax contours overlap for the preferred solution 
($s<1$, $u_0<0$).
This is a useful check on systematics for the case of that event.  However,
because the {\it Spitzer}-``only'' arc and the ground-only ellipse are
essentially tangent at the point of intersection, this is not a case of
combining two 1-D measurements to form a 2-D measurement.
}.  In two of the previous cases, 
KMT-2018-BLG-0029 \citep{kb180029} and OGLE-2018-BLG-0799 \citep{ob180799}
$\bpi_\e$ was basically determined by {\it Spitzer}-``only'', while the much
weaker ground-only information served mainly to help distinguish
among degenerate solutions.  In the other case, Kojima-1 \citep{kojima2},
the ground-only microlensing data also provided relatively weak constraints
that mainly helped distinguish between degenerate solutions.  However,
in this case, there was very precise, purely 1-D information from
VLTI GRAVITY interferometry \citep{kojima1}.  
\\
\par
It would be of interest to determine whether there are other such
cases, in part to determine whether (as for OGLE-2017-BLG-0406)
the {\it Spitzer}-``only'' and ground-only contours were consistent
at the $1\,\sigma$ (or perhaps $2\,\sigma$) level.  This could
provide important statistical information on the frequency of systematic
errors in both types of data sets.  We note that the analytic 
Equations~(\ref{eqn:linear})--(\ref{eqn:linear3}) provide a fast
route to mapping {\it Spitzer}-``only'' contours out to arbitrarily
large $\sigma$.
\\
\par

\subsection{{Future Imaging With Adaptive Optics}
\label{sec:future}}
With its measured mass $M=0.56\pm 0.07\,M_\odot$ and distance $D_{\rm L}=5.2\pm 0.5\,\kpc$,
the OGLE-2017-BLG-0406 host is likely to be $\sim 90$ times fainter
than the microlensed source in the $H$ band, with a $2\sigma$ range of 
31--210 times fainter. According
to Table~\ref{tab:phys}, the lens and source are separating 
at $\mu_{\rm rel,H}=6.1\,\masyr$,
i.e., just slightly larger than the geocentric proper motion. This implies a separation
of $\sim 49\,$mas in 2025 and $\sim 73\,$mas in 2029. With current instrumentation,
the Keck AO system can probably detect the lens in 2029, but with improved instrumentation
\citep{wiz19}, excellent AO corrections may be possible in the $H$ or $J$ bands, allowing
detection in 2025. With an ELT, it may be possible to detect it sooner.
JWST may be able to detect it earlier than 2025 via image elongation \citep{ben07} or the 
color dependent centroid method \citep{ben06,bha19}.
Multi-orbit HST observations might be able to detect the lens by 2025, 
but the source and lens are much too faint for VLT GRAVITY.

\acknowledgments 
Work by Y.H. was supported by JSPS KAKENHI Grant Number 17J02146. 
DPB, AB, and CR  were supported by NASA through grant NASA-80NSSC18K0274.
Work by N.K. is supported by JSPS KAKENHI Grant Number JP18J00897.
Work by CR was supported by an appointment to the NASA Postdoctoral Program
at the Goddard Space Flight Center, administered by USRA through a contract
with NASA.
Work by AG was supported by AST-1516842 from the US NSF
and by JPL grant 1500811.
AG received support from the European  Research  Council  under  the  European  Union’s Seventh Framework Programme (FP 7) ERC Grant Agreement n. [321035]
Work by C.H. was supported by the grants of the National Research Foundation of Korea (2017R1A4A1015178 and 2019R1A2C2085965).
The MOA project is supported by JSPS KAKENHI Grant Number JSPS24253004, JSPS26247023, JSPS23340064, JSPS15H00781, JP16H06287, 17H02871 and 19KK0082 .
The OGLE project has received funding from the National Science Centre,
Poland, grant MAESTRO 2014/14/A/ST9/00121 to AU.
This research has made use of the KMTNet system operated by the Korea
Astronomy and Space Science Institute (KASI) and the data were obtained at
three host sites of CTIO in Chile, SAAO in South Africa, and SSO in
Australia.
YT acknowledges the support of DFG priority program SPP 1992 "Exploring the Diversity of Extrasolar Planets" (WA 1047/11-1).

\appendix
\section{Investigation of Trends in the {\it Spitzer} Residuals}
\label{sec:append}
The {\it Spitzer} residuals to the upper panel of Figure~\ref{fig:lc}
show a low-amplitude (compared to the overall 2017-2019 flux variation)
``wave'' that appears to be driven by six systematically
high points at the end of the 2017 data.  If we remove these six points
from the fit, then $\chi^2$ for the {\it Spitzer} data improves by 
$\Delta\chi^2 =13.5$.  We find numerically that for a Gaussian series 
of 28 points, the probability that there are six consecutive points with 
total contribution $\chi^2>13.5$, and all above or all below the model,
is $1.7\%$.  For a test made a priori, this
would be fairly compelling evidence of systematic effects.  For an a posteriori
test that is constructed to match visually identified features in the data, it
is less so.  Nevertheless, this test motivates us to check the impact on the
final results of including versus excluding these six points.
\\
\par
\begin{figure}
\plotone{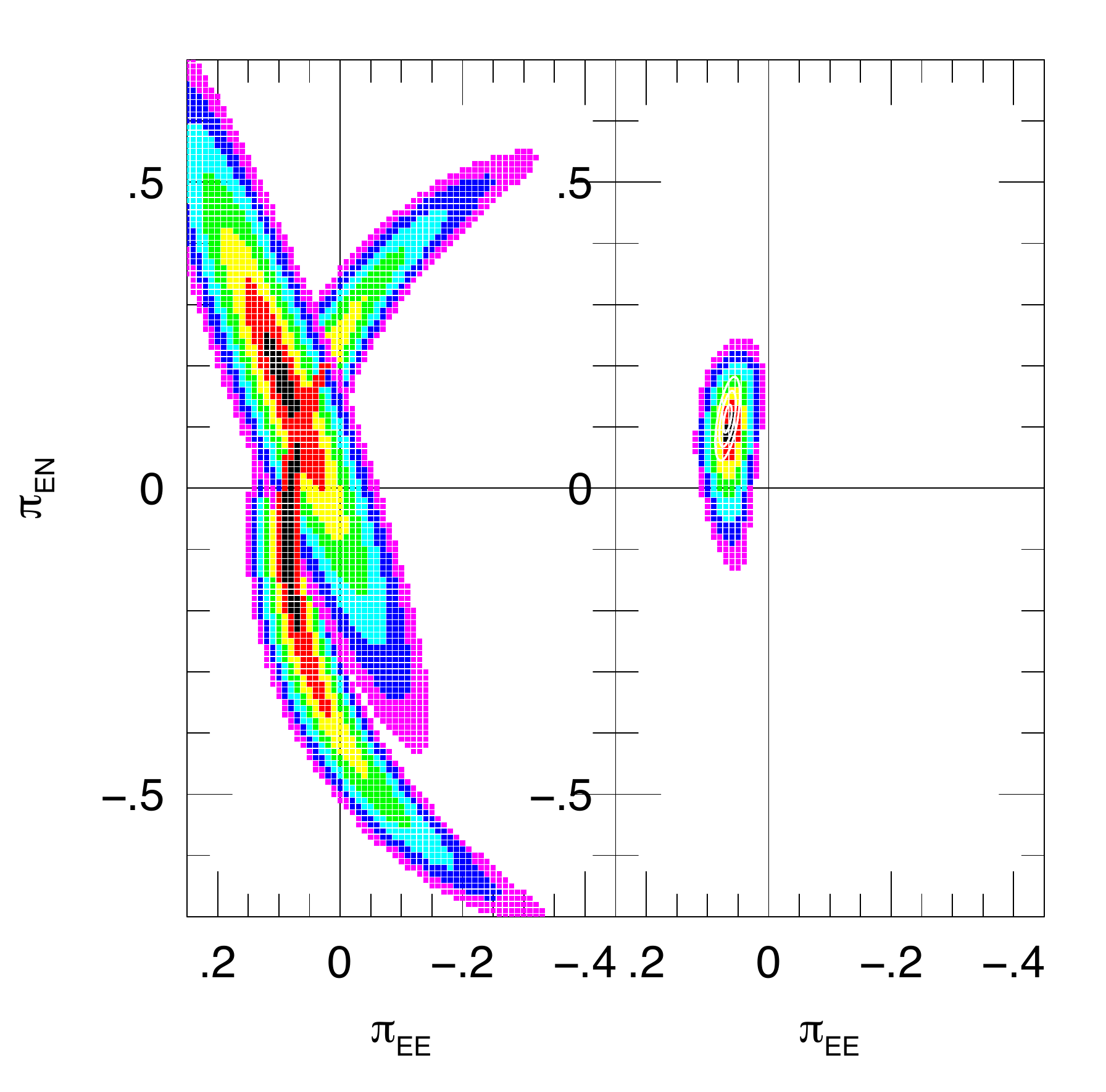}
\caption{OGLE-2017-BLG-0406 parallax contours for the W+ solution
after the removal of the last six {\it Spitzer} data points from 2017.  
Hence, the elliptical contours from the ground-only fit in the left panel
are identical to those of Figure~\ref{fig:arc+}.  The {\it Spitzer}
arc looks qualitatively similar, but there is a single minimum
near $\pi_{\e,N}\sim 0$ rather than two weak, roughly symmetric 
minima at $\pm\pi_{\e,N}$.  Nevertheless, in both cases the two
sets of contours overlap at $1\,\sigma$.  The colored
contours at the right show the product of the two sets of 
likelihood contours from the left.  The white contours are from the
full fit to all the data, i.e. they are the same as in 
Figure~\ref{fig:arc+}.  Thus comparison of the white and
colored contours shows that the solution changes by $<1\,\sigma$.
}
\label{fig:arc+rm}
\end{figure}

We therefore repeat the analysis of Section~\ref{sec:spitz-joint} after
first removing the six points and also re-renormalize the error bars by a
further factor of 0.78 in order to enforce $\chi^2/{\rm dof}=1$.  
Figure~\ref{fig:arc+rm} 
shows the resulting analog of Figure~\ref{fig:arc+}.
The main change is that the {\it Spitzer}-``only'' minimum 
slides ``along the arc''
toward its center.  However, the point of intersection with the ground-only
solution barely changes, with the net result being that the combined solution
(right panel) barely changes.
\\
\par

This relative insensitivity of the final result to the details of the late-time
2017 data can be understood within the framework of the \citet{gould19}
``osculating circle'' analysis.  Each of the earliest 2017 points 
(together with the 2019 baseline and the color-color constraint) yields a
circle in the $\bpi_\e$ plane.  These osculating circles differ slightly in
radius (and center), and so their overlap produces an arc.  
Late-time data can fill in
the details of this arc, but these details are largely irrelevant because,
to zeroth order, it is only the point where the arc crosses the major axis of 
the ground-only error ellipse that defines the solution.  At first order, the
position of the minimum along the {\it Spitzer}-``only'' arc plays some role 
because the ground-only contours
have a finite width.  However, as this width is small, the best fit can only 
be moved from one side to the other of this narrow range.  
\\
\par
Tables~\ref{tab:combined_rem}
 and \ref{tab:phys_rem} give the microlensing and physical parameters for the
W+ and W- solutions.  They can be directly compared to 
Tables~\ref{tab:combined} and \ref{tab:phys}.  As would be predicted from the
analysis of 
Figure~\ref{fig:arc+rm}, 
the physical parameters change by
very little:  $\leq 0.5\,\sigma$ for the masses and by less than half that
for the distances and velocities.  
\\
\par

\begin{deluxetable}{lcc}
\tablecolumns{3} \tablewidth{0pc} \tablecaption{\textsc{Wide models
for ground+{\it Spitzer} data with 6 pts removed}} \tablehead{
\colhead{Parameters } & \colhead{Wide$(+,+)$} &
\colhead{Wide$(-,+)$} } \startdata
  $t_0$ $(\rm{HJD}^{\prime})$     &7908.812 $\pm$ 0.001     &7908.813 $\pm$ 0.001  \\
  $u_0$ $(10^{-3})$               &9.298 $\pm$ 0.028        &-9.281 $\pm$ 0.028    \\
  $t_{\rm E}$ $(\rm{days})$       &37.069 $\pm$ 0.086       &37.133 $\pm$ 0.084    \\
  $s$                             &1.128 $\pm$ 0.001        &1.128 $\pm$ 0.001     \\
  $q$ $(10^{-4})$                 &6.924 $\pm$ 0.091        &6.970 $\pm$ 0.090     \\
  $\alpha$ $(\rm{rad})$           &0.993 $\pm$ 0.001        &-0.993 $\pm$ 0.002    \\
  $\rho$ $(10^{-3})$              &5.861 $\pm$ 0.024        &5.843 $\pm$ 0.025     \\
  $\pi_{\rm{E},\it{N}}$           &0.117(0.101) $\pm$ 0.023 &0.120(0.105) $\pm$ 0.024  \\
  $\pi_{\rm{E},\it{E}}$           &0.063(0.066) $\pm$ 0.007 &0.065(0.067) $\pm$ 0.007  \\
  $\pi_{\rm{E}}$                  &0.133(0.121) $\pm$ 0.017 &0.136(0.126) $\pm$ 0.018  \\
  $\phi_\pi$                      &0.495(0.593) $\pm$ 0.143 &0.499(0.584) $\pm$ 0.141  \\
  $f_S({\rm OGLE})$               &1.461 $\pm$ 0.004        &1.459 $\pm$ 0.004     \\
  $f_B({\rm OGLE})$               &0.103 $\pm$ 0.004        &0.105 $\pm$ 0.004     \\
  $f_S(Spitzer)$                  &11.297 $\pm$ 0.175       &11.265 $\pm$ 0.182    \\
  $f_B(Spitzer)$                  &-2.841 $\pm$ 0.176       &-2.808 $\pm$ 0.183    \\
  $t_*$ $(\rm{days})$             &0.217 $\pm$ 0.001        &0.217 $\pm$ 0.001     \\
\enddata
\tablecomments{Mean values from the MCMC are shown in parentheses.
All other values are from the best-fit model. $\pi_\e$, $\phi_\pi$,
and $t_*$ are derived quantities and are not fitted independently.
All fluxes are on an 18th magnitude scale, e.g., $I_s=
18-2.5\,\log(f_s)$.} \label{tab:combined_rem}
\end{deluxetable}

\begin{deluxetable}{lcr} 
\tablecolumns{3} \tablewidth{0pc}
\tablecaption{\textsc{Physical Parameters\label{tab:phys_rem}}}
\tablehead{ \colhead{Parameter} & \colhead{units} &
\colhead{Values}}
\startdata
$M_{\rm host}$  & $M_\odot$         & $  0.60\pm  0.09$ \\
$M_{\rm planet}$  & $M_{\rm Jup}$   & $  0.44\pm  0.06$ \\
$D_{\rm L}$  & kpc                        & $  5.10\pm  0.39$\\
$\mu_{\rm rel,H,N}$ & mas/yr            & $  4.82\pm  0.52$\\
$\mu_{\rm rel,H,E}$    & mas/yr             & $  3.64\pm  0.61$\\
$a_\perp$  & au                    & $  3.40\pm  0.26$\\
\enddata
\end{deluxetable}

The causes of the systematic trends in the {\it Spitzer} data are
not fully understood.  \citet{kb180029} found trends of a generally similar form, but of
much greater amplitude relative to the observed {\it Spitzer} flux variation,
in their analysis of KMT-2018-BLG-0029.  They argued that these were most likely
due to the effect of normal field rotation during the {\it Spitzer} observing
window combined with the poorly determined positions of several nearby stars
that were many magnitudes brighter than the source.  They therefore argued that
the most robust delta-flux measurements in the light curve were those
between observations early in the 2018 and early in the 2019 seasons,
which all had similar field angle.
\\
\par
In the present case, there are no such bright contaminating nearby stars.
And correspondingly, the observed trends are much weaker.  Hence, perhaps
there is some similar effect from the wings of more distant stars.
In any case, the net effect of eliminating the last six points from 2017
is that all remaining data are from the beginning of the observing window
(first 9.3 days of 2017 and first 7.0 days of 2019), when we expect the 
effects of field rotation to be minimized.  
\\
\par
We infer that the radius of the osculating 
circles, which is the aspect of the {\it Spitzer} parallax measurement
that primarily contributes to the final $\bpi_\e$ measurement, is the least 
subject to systematic effects, because it derives directly from the 
comparison of early-2017 with early-2019 data, which are at the same field
orientation.  That is, the narrow (best statistically
determined) direction of the {\it Spitzer} contours, namely, the radial 
coordinate defined by the circular arc,  is also the most
robust from the standpoint of systematics.  Recall from 
Section~\ref{sec:spitz-joint} that the same was true of the ground-based
contours: the narrow (best statistically determined) direction was also
the more robust from the standpoint of systematic errors.
\\
\par
Because the case for removing the final six points from 2017 is not
compelling and also because doing so changes the estimates of
the physical parameters by substantially less than $1\,\sigma$, we report the
determinations from the full data set as our results.  However, for 
completeness, we also list the results from fits with these six points
removed in Tables~\ref{tab:combined_rem} and \ref{tab:phys_rem}.
\\
\par

\end{document}